\documentclass[fleqn,usenatbib]{mnras}

\usepackage{newtxtext,newtxmath}

\usepackage[T1]{fontenc}

\DeclareRobustCommand{\VAN}[3]{#2}
\let\VANthebibliography\thebibliography
\def\thebibliography{\DeclareRobustCommand{\VAN}[3]{##3}\VANthebibliography}

\usepackage{graphicx}	
\usepackage{amsmath}	

\usepackage{amssymb}	

\usepackage{enumitem}
\usepackage{bm}
\usepackage{multirow}
\usepackage{lscape} 
\graphicspath{{plots/}}
\usepackage{pdflscape}
\usepackage{xcolor,colortbl}
\usepackage{color}
\usepackage{textcomp}
\usepackage[normalem]{ulem}

\usepackage{xspace}
\newcommand{\gadgetx}{\textsc{Gadget-X}\xspace}
\newcommand{\ahf}{\textsc{AHF}\xspace}
\newcommand{\oPDF}{\textsc{oPDF}\xspace}
\newcommand{\OOB}{\textsc{OOB}\xspace}
\newcommand{\RFR}{\textsc{RFR}\xspace}
\newcommand{\DS}{\textsc{DS}\xspace}
\newcommand{\ud}{\mathrm{d}}
\newcommand{\mathL}{\mathcal{L}}
\newcommand{\deltaL}{$\Delta \ln \mathL$ }

\usepackage{natbib}
\defcitealias{paperI}{Paper~I}

\usepackage{newtxtext,newtxmath}

\title[Dynamical state]{What to expect from dynamical modelling of cluster haloes II. Investigating dynamical state indicators with Random Forest}
\author[Li et al.]{\parbox{\textwidth}{
Qingyang Li,$^{1,2,3}$\thanks{E-mail: qingyli@sjtu.edu.cn}
Jiaxin Han,$^{1,2,3}$\thanks{E-mail: jiaxin.han@sjtu.edu.cn}
Wenting Wang,$^{1,2,3}$\thanks{E-mail: wenting.wang@sjtu.edu.cn}
Weiguang Cui,$^{4}$
Federico De Luca,$^{5}$
Xiaohu Yang,$^{1,2,3,6}$
Yanrui Zhou,$^{1,2,3}$ and
Rui Shi$^{1,2,3}$
}
\vspace{0.4cm}
\\
\parbox{\textwidth}{
$^{1}$Department of Astronomy, School of Physics and Astronomy, Shanghai Jiao Tong University, Shanghai 200240, China\\
$^{2}$ Key Laboratory for Particle Astrophysics and Cosmology (MOE), Shanghai 200240, China\\
$^{3}$ Shanghai Key Laboratory for Particle Physics and Cosmology, Shanghai 200240, China\\
$^{4}$Institute for Astronomy, University of Edinburgh, Royal Observatory, Edinburgh EH9 3HJ, United Kingdom\\
$^{5}$Dipartimento di Fisica, Universit$\grave{a}$ di Roma “Tor Vergata”, Via della Ricerca Scientifica 1, 00133 Roma, Italy\\
$^{6}$Tsung-Dao Lee Institute, Shanghai Jiao Tong University, Shanghai 200240, China
}}

\date{Accepted XXX. Received YYY; in original form ZZZ}

\pubyear{2021}

\begin{document}
\label{firstpage}
\pagerange{\pageref{firstpage}--\pageref{lastpage}}
\maketitle

\begin{abstract}


We investigate the importances of various dynamical features in predicting the dynamical state (\DS) of galaxy clusters, based on the Random Forest (RF) machine learning approach. We use a large sample of galaxy clusters from the Three Hundred Project of hydrodynamical zoomed-in simulations, and construct dynamical features from the raw data as well as from the corresponding mock maps in the optical, X-ray, and Sunyaev–Zel’dovich (SZ) channels. Instead of relying on the impurity based feature importance of the RF algorithm, we directly use the out-of-bag (\OOB) scores to evaluate the importances of individual features and different feature combinations. Among all the features studied, we find the virial ratio, $\eta$, to be the most important single feature. The features calculated directly from the simulations and in 3-dimensions carry more information on the \DS than those constructed from the mock maps. Compared with the features based on X-ray or SZ maps, features related to the centroid positions are more important. Despite the large number of investigated features, a combination of up to three features of different types can already saturate the score of the prediction. Lastly, we show that the most sensitive feature $\eta$ is strongly correlated with the well-known half-mass bias in dynamical modelling. Without a selection in \DS, cluster halos have an asymmetric distribution in $\eta$, corresponding to an overall positive half-mass bias. Our work provides a quantitative reference for selecting the best features to discriminate the \DS of galaxy clusters in both simulations and observations.
\end{abstract}




\section{Introduction}

Galaxy clusters, which are embedded in the most massive populations of dark matter haloes and contribute to the most luminous end of galaxy distribution in our Universe, are crucial objects to study in the field of galaxy formation and cosmology  \citep[e.g.][]{Yang2007, Rykoff2014, Yang2021}. They provide suitable environments to examine the quenching of star formation in both the central massive galaxies and other smaller member satellite galaxies \citep[e.g.][]{Kimm2009, Wetzel2013,Boselli2016,Wang2018, PC2019}, to look for missing baryons \citep[e.g.][]{Monteagudo2015,Graaff2019, Lim2020}, to investigate the hot gas distribution through X-ray and Sunyaev-Zeldovich (SZ) observations \citep[e.g.][]{Arnaud2010,Planck2013, Lim2018}, to study the connection between galaxies, hot gas and the host dark matter haloes \citep[e.g.][]{Planck2013,Anderson2015,Wang2016} and even serve as possible standard rulers in cosmology \citep[e.g.][]{Wagoner2021}. In the era of precision cosmology, accurate determination of the total mass of galaxy clusters, which is dominated by the invisible dark matter halo, is a very important prerequisite for robust scientific conclusions in these different fields. In addition to weak gravitational lensing \citep[e.g.][]{Han15, Luo2018, Sun2021}, the total mass of galaxy clusters is often measured through dynamical modelling of observed tracer objects, such as the distribution of hot gas and satellite galaxies in clusters \citep[e.g.,][]{Diaferio1997,Biviano2006,Wojtak2008,Rasia2012,Mamon2013,Old2014,Ramanah2021,paperI}.

Most dynamical models, such as the Jeans equation \citep[e.g.,][]{Binney1987} and hydrostatic equilibrium equation \citep[e.g.,][]{Rasia2004}, have to adopt the steady-state and spherical assumptions. The validity of these assumptions depends on the dynamical state (hereafter \DS) of real galaxy clusters. The \DS of galaxy clusters has been studied in both simulations and observations. Different features have been defined to separate relaxed and unrelaxed clusters. For example, \citet{Bett2007} used the virial ratio parameter, which is the ratio between the kinetic and potential energy, to select quasi-equilibrium haloes. \citet{Neto2007} chose the fraction of mass in substructures and the offset between the centre of mass and the potential minimum to quantify the \DS. These features are revisited and combined with each other to describe the \DS of galaxy clusters in many later studies  \citep[e.g.,][]{Knebe2008,Davis2011,Ludlow2012,Power2012,Cui2017,Cui2018,DeLuca2021,Capalbo2021}.

Observationally, the most popular features to quantify the \DS of galaxy clusters are often based on X-ray images or SZ maps such as the galaxy concentration \citep[e.g.,][]{Santos2008,Cassano2010}, the offset between centroids defined in different ways \citep[e.g.,][]{Mohr1993,Maughan2008} and the power ratio \citep[e.g.,][see Section~\ref{sec:features} for details]{Buote1995,Bohringer2010}. Some studies have adopted a weighted mixture of different features \citep[e.g.,][]{Mann2012,Zenteno2020}. For example, \citet{Yuan2020} defined a morphology index, in combination with a parameter defined through the surface brightness profile and an asymmetry factor based on images from the \emph{Chandra} X-ray observation, to quantify the \DS of galaxy clusters. In addition, optical data can provide dynamical information \citep[e.g.,][]{Ribeiro2013} as well. For instance, by using photometric data from the Sloan Digital Sky Survey (SDSS), \citet{Wen2013} developed a relaxation parameter to quantify the \DS, by taking into account the asymmetry, the ridge flatness and the normalized pixel differences between the smoothed optical map and the best-fitting elliptical King model. The offsets between the intensity peaks in different bands are also used to indicate the \DS \citep[e.g.][]{Mann2012,Yasuhiro2014,Rossetti2016,Zenteno2020}. More recently, correlations between the \DS and the fraction of stellar mass in the intracluster light (ICL) have also been reported \citep[e.g.,][]{Yolanda2018}, though still under debates. 

Although many different features have been developed, comprehensively understanding the \DS of galaxy clusters is not easy with a single feature. Besides, it is important to determine which feature is a more effective proxy to the \DS and how many features are required to achieve a good description of the \DS. These still remain uncertain in previous studies. \citet{Cialone2018} showed the stability and efficiency of morphological parameters from mock SZ maps in classifying the dynamical status of galaxy clusters, and claimed some weak correlations between these combined parameters and the \DS. \citet{DeLuca2021} evaluated the efficiency of morphological parameters constructed from mock observations using the Kolmogorov-Smirnov (KS) test and the Receiver Operating Characteristic (ROC) curve. They found that the combined morphological parameters provide more information about the \DS of galaxy clusters. 

Dynamical modelling provides a unique way to answer these questions, and numerical simulations provide us with the ground truth to be compared with the best fits through dynamical modelling. Galaxy clusters deviating from the steady-state assumption are expected to have larger
biases and scatters in the best-fitting versus true virial masses of the host dark matter halo. For example, based on the orbital Probability Distribution Function \citep[oPDF,][]{Han2016a} and the spherical Jeans equation, \cite{paperI} (hereafter \citetalias{paperI}) showed that the best-fitting halo mass and concentration for relaxed clusters have smaller scatters than unrelaxed clusters. In \citetalias{paperI}, clusters are divided into relaxed and unrelaxed subsamples according to the combined criteria based on the virial ratio, the fraction of mass in subhaloes and the offset between the mass center and the density peak \citep{Cui2018}. In this study, we generalize the analysis to use a much larger feature set, and study what features are most responsible for the amount of systematics in dynamical modelling. 

The goal of this study is to identify efficient dynamical indicators that can tell the dynamical state of a cluster. Theoretically, these dynamical indicators could help us to pin down the most relevant physical processes responsible for the deviation from equilibrium. Practically, these indicators may also serve as continuous tags to label the level of systematic biases in dynamical modelling, beyond classifying them into relaxed and un-relaxed classes.

To this end, we adopt the log-likelihood difference of the \oPDF model evaluated at the best-fitting and true halo parameters as a unique representation of the \DS of galaxy clusters. We then investigate the importances of different cluster features in determining the log-likelihood difference and hence the \DS using a machine learning approach, the Random Forest (RF) method \citep{Breiman2001}. As a relatively new and powerful statistical analysis tool, machine learning is popularly applied to many different aspects of galaxy cluster studies in recent years, such as cluster mass estimation \citep[e.g.,][]{Ntampaka2015,Ntampaka2016,Armitage2019,Ntampaka2019,Green2019,Gupta2020,Kodi2020,Cohn2020,Yan2020,Ho2021}, census of galaxy clusters \citep{Su2020}, determination of cluster sample \citep[e.g.,][]{Lin2021,Angora2020} and SZ effect \citep[e.g,][]{Bonjean2020,Rothschild2022,Wadekar2022}. In particular, there are growing interests of using machine learning to improve cluster mass estimates. For example, with convolutional neural networks, \citet{Ho2019} achieved a lognormal residual of scatter as low as 0.132 dex for cluster mass prediction.

Among the various kinds of machine learning approaches, the RF method is one of the most effective and robust machine learning algorithms. It has been applied in many different fields of astrophysics, such as the redshift-stellar mass distribution \citep[e.g.][]{Mucesh2021}, halo occupation and galaxy assembly bias \citep[e.g.][]{Xu2021}, photometric redshifts \citep[e.g.][]{Carliles2010,CB2013}, gamma/hadron separation \citep[e.g.][]{Albert2008}, estimates of cluster and group mass \citep[e.g.][]{Green2019,Man2019}, prediction of accreted stellar mass fractions \citep[][]{Shi2021} and the dependence of halo mass on central-satellite magnitude gaps \citep[][]{Zhou2022}. The RF method is widely used because of its simplicity, accuracy and efficiency. It is capable of fitting a nonlinear model to a high dimensional dataset non-parametrically, while providing an objective way to evaluate the goodness of fit. More importantly, the RF method comes with metrics to evaluate the importances of different features. This enables us to rank and disentangle the contributions from different features to the prediction, which is crucial for our analysis of selecting \DS indicators. 

This paper is organised as follows. In Section~\ref{sec:data}, we briefly introduce the simulation data and mock X-ray and SZ maps. We introduce the RF method, the target variable, features of galaxy clusters and other details in Section~\ref{sec:method}. We present general results about the RF learning outcome, the importance ranking of different features and the best out-of-bag (\OOB) score we can achieve in Section~\ref{sec:performance}. In Section~\ref{sec:disc}, we discuss how the bias in best-fitting halo parameters depend on the virial ratio and the mass enclosed within the half-mass radius. The conclusions are summarised in Section~\ref{sec:con}.

\section{Data} \label{sec:data}

\subsection{Galaxy clusters in the Three Hundred Project}

Our sample of galaxy clusters are based on the suite of hydrodynamical simulations from the Three Hundred Project\footnote{\url{https://the300-project.org}}~\citep{Cui2018}. The simulations are performed with \gadgetx, one of the state-of-the-art hydrodynamical codes.
The \gadgetx model includes black hole (BH) growth and implementation of AGN feedback \citep{Steinborn2015}, in addition to the basic sub-grid models. The dark matter and gas particle masses in the high resolution region are $m_{\rm DM}\simeq 12.7 \times 10^8\ h^{-1}\mathrm{M_{\odot}}$ and $m_{\rm gas} \simeq 2.36\ \times 10^8\ h^{-1}\mathrm{M_{\odot}}$, respectively. 
There are 324 clusters in total, which are selected as those systems which have host halo masses $M_{\rm h}\gtrsim 8 \times 10^{14}\ h^{-1}\rm M_{\odot}$ at $z = 0$ and from the parent $N$-body simulation of MultiDark Planck 2~\citep[MDPL2;][]{Klypin2016}\footnote{\url{https://www.cosmosim.org/cms/simulations/mdpl2}}. Here the halo mass, $M_{\rm h}$, is defined as the mass enclosed inside an overdensity of  $\sim$98 times the critical density of the universe \citep{Bryan1998}. These clusters are re-simulated with \gadgetx\ under the same Planck cosmology as the parent simulation~\citep{Planck2016}. 

Galaxy clusters in the Three Hundred Project have been used to study many different aspects of galaxy formation and cosmology, including, for example, the environmental effects \citep{WangYang2018}, cluster profiles \citep{Mostoghiu2019,Li2020,Baxter2021}, splash-back galaxies \citep{Arthur2019,Haggar2020,Knebe2020}, cluster dynamical state \citep{DeLuca2021,Capalbo2021,Zhang2021}, filament structures \citep{Kuchner2020,Rost2021,Kuchner2021} and gravitational lensing \citep{Vega-Ferrero2021}. Readers can find more details in \citet{Cui2018} and \citetalias{paperI}. 

Dark matter haloes and substrctures in these re-simulations are identified by the Amiga Halo Finder~\citep[\ahf;][]{Knollmann2009}. In our analysis, the true mass for each halo is defined as the virial mass ($M_{200}$) as identified by \ahf\footnote{The virial mass, $M_{200}$, is defined as the total mass enclosed in the radius, $r_{200}$, within which the average matter density is 200 times the critical density, $\rm \rho_{crit}$, of the universe.}. The true halo concentration is defined as $r_{200}/r_\mathrm{s}$, where $r_\mathrm{s}$ is the scale radius obtained by fitting a two-parameter Navarro-Frenk-White~\citep[NFW;][]{NFW} model profile to the actual radial density profile of the total mass distribution.\footnote{Note that our template fitting approach allows us to choose any definition of the scale radius and concentration parameters, and an unbiased fit will always recover the true parameters by construction~\citep[see Appendix A of][]{Han2016b}.} 

In our sample, we select the clusters at $z=0$ without catastrophic failures during the dynamical modelling. Specifically, we remove the clusters whose best-fitting parameters lie outside the 3$\sigma$ confidence region of the sample distribution. This is done iteratively until the remaining number of clusters converge. Finally, we have 310 clusters in total with mass $M_{200} > 5.21 \times 10^{14}\ h^{-1}\rm M_{\odot}$. 

\subsection{Mock maps}

For the sample of galaxy clusters selected from the Three Hundred Project above, mock optical, X-ray and SZ maps are created. The mock maps are projected along the $z$-direction and centred on the density/intensity peaks, which cover circular regions with the radius of 1.4$r_{200}$. The $z=0$ clusters are placed at a redshift of $z=0.05$ to create the mock observations. The contaminations by the sky background and instrumental noise are not included. More details about the mock maps are presented in \citet{DeLuca2021}. Here we just make a brief introduction.

The optical maps are created for the SDSS $r$-band filter, with a pixel size of 0.396$\arcsec$. The surface brightness distributions are derived through the stellar population synthesis modelling \citep[see details in][]{Cui2011,Cui2014,Cui2016}, with the spectrum of each star particle produced by interpolating the \citet{BC2003} stellar evolution library.

The X-ray emission of galaxy clusters is modelled as bremsstrahlung emission from the hot electrons in the intracluster medium (ICM). The X-ray images are produced by applying the \textsc{py}XSIM code \citep{ZuHone2014,ZuHone2016}, with 10~ks of exposure time and a bandpass of 0.1-15 $\rm keV$. The number of electrons per gas particle, $N_i$, is calculated with a metallicity dependence (more details can be found in \citep{Cui2018}): \begin{equation}
    N_{i} = \frac{N_{e} m_i (1-Z-Y_{\rm He})}{\mu  m_\mathrm{p}}.
\end{equation}
In the equation, $N_{e}$ is the number of ionized electrons per hydrogen particle. $m_i$ is the mass of the gas particle. $Z$ is the metallicity of the gas particle. $Y_{\rm He}$ is the helium mass fraction of the gas particle. $\mu$ is the mean molecular weight, and $m_\mathrm{p}$ is the proton mass. The electron number density is computed by using all gas particles without star formations, but we did not apply any temperature cut. This is because the fraction of cold gas in galaxy clusters is very small \citep{Li2020}, and not applying temperature cuts barely affects the results.
The bolometric flux of photons received by the observer for a given volume element $\Delta V_i$ is:
\begin{equation}
    F_{i}^{\gamma} = \frac{n_{e,i} n_{H,i} \Lambda(T_i,Z_i)\Delta V_i}{4\pi D^2_{A,0}(1+z_0)^2}\ \rm photons\ s^{-1}\ cm^{-2},
\end{equation}
where the superscript $\gamma$ means it is a photon count emissivity. $z_0$ is the redshift of the source. $D_{A,0}$ is the angular diameter distance to the source. $\Lambda(T_i,Z_i)$ is the spectral model as a function of temperature ($T_i$) and metallicity ($Z_i$). $n_e$ and $n_H$ are number densities of electrons and protons, respectively.

The SZ effect is produced from the inverse Compton scattering of the Cosmic Microwave Background (CMB) photons by hot electrons in the ICM. There are two different types of SZ effect, thermal and kinematic. Here we only focus on the thermal SZ effect, which is usually more prominent than the kinetic SZ effect. The thermal SZ maps are produced with the \textsc{py}MSZ code\footnote{\url{https://github.com/weiguangcui/pymsz}} based on a discrete version of the dimensionless Comptonization parameter $y$:
\begin{equation}
    y = \frac{\sigma_T k}{m_e c^2 dA} \sum_i T_i N_{e,i} W(r,h_i),
\end{equation}
where $m_e, T_i$ and $N_{e,i}$ are the mass, temperature and number of electrons, respectively. $\sigma_T$ is the Thomson cross section, $k$ is the Boltzmann constant, $c$ is the light speed, $dA$ is the pixel area and $W(r,h_i)$ is the smoothed particle hydrodynamics (SPH) smoothing kernel. The summation is over all the gas particles in pixel $i$. The SZ signals are usually demonstrated by the map of this $y$ Compton parameter, which we will simply call it SZ maps throughout this paper.

The spatial resolution of ICM maps is fixed to 10 kpc $\rm pixel^{-1}$ (comoving). We will introduce the morphological features constructed from these maps in Section~\ref{sec:features}.

\section{Method} \label{sec:method}

\subsection{Dynamical method}\label{sec:oPDF}
We use a generic dynamical model -- the orbital Probability Density Function (\oPDF\footnote{Code available at \url{https://github.com/Kambrian/oPDF}}) which involves only the most commonly used assumptions: the steady-state and spherical symmetry. Deviations from these assumptions are also the defining characteristics of the \DS. This estimator is a freeform distribution function method that works by fitting a data-driven distribution function to the tracer sample to infer the underlying potential \citep{Han2016a}. In this work, the potential profile, $\Phi(r)$, is parametrized by generalizing the true potential to a template with the formula $\Phi(r) = A \Phi_{\rm true}(Br)$, where the true profile $\Phi_{\rm true}(r)$ is extracted from the simulation, and $A$ and $B$ are free parameters which can be converted to mass and concentration of clusters following \citet{Han2016b}.

For a steady-state spatial tracer particle, the prediction of distribution relies on the \oPDF. If a system is in a steady-state, then along a given orbit the probability of observing a particle near any position (labelled by the parameter $\lambda$) is proportional to the time that the particle spends around that position, i.e.:
\begin{equation}
    \frac{\ud P(\lambda|{orbit})}{\ud\lambda} \propto \frac{\ud t(\lambda|{orbit})}{\ud\lambda}.
\end{equation} 
This conditional distribution can be derived from the time-independent collisionless Boltzmann equation and is equivalent to the Jeans theorem~\citep{Han2016a}. Applied to a spherical system, the overall steady-state radial distribution of all the particles is then predicted combining the \oPDF of each particle as
\begin{equation}
    P(r)=\frac{1}{N} \Sigma_i P(r|E_i,L_i),\label{eq:P_r}
\end{equation}
where $E_i$ and $L_i$ are the binding energy and angular momentum of each tracer particle respectively. The underlying potential can be inferred with this predicted distribution function, by matching the predicted distribution with the observed distribution in a statistical framework. 

Combined with the binned likelihood approach for the inference, the likelihood of observing the data with the model distribution of Equation~\eqref{eq:P_r} can be written as 
\begin{equation}
    \begin{split}
        \mathL &=\bm{\prod}_{j=1}^m \hat{n}_j^{n_j} \exp(-\hat{n}_j) \\
        &=\exp(-N) \bm{\prod}_{j=1}^m \hat{n}_j^{n_j},
    \end{split}
\end{equation}
where $n_j$ and $\hat{n}_j$ are the the observed and predicted numbers of particles in the $j$-th bin respectively. The best-fitting potential can be inferred by searching for a potential that maximizes this likelihood. 

If the halo is spherical and in a steady-state, then the best fitting parameters, $(\hat{M}, \hat{c})$, are expected to be close to the true parameters, $(M_{\rm true}, c_{\rm true})$, up to the statistical uncertainties. Correspondingly, the likelihood function evaluated at the best fitting and the true parameter values will also be close to each other. More precisely, according to Wilks's theorem~\citep{Wilks1938}, the double log-likelihood ratio,
\begin{equation}
     2\Delta \ln \mathL = 2[\ln \mathL (\hat{M}, \hat{c})-\ln \mathL(M_{\rm true}, c_{\rm true})],
\end{equation} is expected to be a $\chi^2$ variable with two degrees of freedom in this case, regardless of the statistical uncertainties. 

In reality, a halo is neither perfectly spherical or in a complete steady-state. These deviations will lead to larger deviations in the best-fitting parameters and the corresponding likelihood ratios. As a result, the likelihood ratio can quantify the amount of systematic deviations in the halo and thus serve as a theoretical \DS measure. Unlike the biases in the best-fitting parameters whose magnitudes are modulated by the statistical uncertainties of the inference, the magnitude of the likelihood ratio is independent of the statistical uncertainty and only determined by the level of systematics. We will discuss this quantity further in section~\ref{sec:target}.

\subsection{Random Forest Regression}
\label{sec:RFRmethod}
RF can be applied to both classification and regression problems. In this paper, we adopt 
the Random Forest Regression \citep[RFR,][]{Breiman2001} to deal with the continuous target variable in our analysis. 
RFR predicts the target variable based on a few different input features. In this work, the target is the log-likelihood ratio between true and best-fitting halo parameters from our dynamically modelling, while the features are various galaxy cluster properties constructed from hydrodynamical simulations or mock maps. The learning outcome of RFR is quantified through the so-called \OOB score. We will introduce in more detail the above concepts in the following sections.

RFR is a bagging algorithm that uses an ensemble of decision trees to generate the final prediction by averaging the output of each tree. This method possesses the ability to deal with large and multi-dimensional datasets and is well suited for studying non-linear problems with a low computation cost. More 
importantly, using the averaged prediction of multiple trees avoids the proneness to over-fitting of an individual decision tree. A whole forest not only preserves the low bias of a single decision tree but also decreases the variance to successfully navigate the bias-variance balance \citep{Mucesh2021}.

The construction of one decision tree starts from the segregation of the training sample according to the 
distribution of input features. Starting from the root node, which is the whole parent sample, it continues splitting the input sample into two subsamples until the sample size reaches a given minimum value on the leaf node. For each node, it iterates over all available features and feature values to decide the best feature, $f_i$, and the best feature value, $\theta_i$, which minimises the mean square error (MSE) after splitting, with the two subsamples divided according to $f_i<\theta_i$ or $f_i>\theta_i$. Other decision trees in the forest are then generated with the same approach but are based on different bootstrap realizations of the training set. 

Once a forest is built up against the training sample, it can be used to predict the target value given a set of input features. This is done by walking the decision tree and find the corresponding leaf node for the input data. The mean value of the corresponding leaf node then forms the prediction for a particular tree, and the final prediction is the average over the predictions of each tree.  

In this work we use the \textsc{RandomForestRegressor} module implemented in the \textsc{python} machine learning package \textsc{scikit-learn} \citep{scikit-learn} for our analysis. The algorithm contains several hyperparameters that can be tuned when building a forest, including \emph{n\_estimators} and \emph{max\_features}. \emph{n\_estimators} represents the number of decision trees used in a forest. Results based on too few trees are not stable. Theoretically, larger values of \emph{n\_estimators} would produce better results, at the cost of a longer computation time. After performing convergence tests, we find the learning outcome barely changes for \emph{n\_estimators}$\sim>$ 400. We thus fix this hyperparameter to 400. \emph{max\_features} is the number of features to be iterated over when looking for the best split during tree construction, which we fix to be the total number of input features. We have tested that a different choice of \emph{max\_features} barely changes our results.

Another important hyperparameter is \emph{min\_samples\_leaf}, which is defined as the minimum sample size of a leaf node. The tree stops splitting if the sample size on a leaf node is smaller than this value. Careful convergence tests show that the best choice of \emph{min\_samples\_leaf} is 5 when all features from the simulation are used. When only features constructed from X-ray or SZ mock maps are used, we found the best choice of \emph{min\_samples\_leaf} is 10. When individual features are used as inputs, the optimal choice of \emph{min\_samples\_leaf} can vary, especially for features with low importances. We thus conduct the convergence test for each individual feature, and determine the best choice of \emph{min\_samples\_leaf} for each feature. 

For our sample of galaxy clusters, we randomly choose a 70\% subsample as our training sample.
In the following, we move on to introduce the target variable and features used in RFR.


\subsubsection{Target}\label{sec:target}

The target variable of \RFR is chosen as the log-likelihood difference, \deltaL, between the best-fitting and true model parameters. It directly reflects the systematic uncertainties \citep{Wang2017} of the model and the \DS of the system. As introduced in section~\ref{sec:oPDF},  we expect our $\Delta \ln{\mathL}$ to behave like a $\chi^2(2)$ variable if the tracers are in steady states and there is no violation of spherical symmetry. From a statistical point of view, \deltaL is a random variable resulting from fitting one random realization of the underlying model. However, such a stochasticity in \deltaL cannot be captured by RFR, which introduces an upper limit to the score of the regression. More detailed discussions about this will be provided in Section~\ref{sec:basic}.

\begin{figure}
    \centering
    \includegraphics[width=0.48\textwidth]{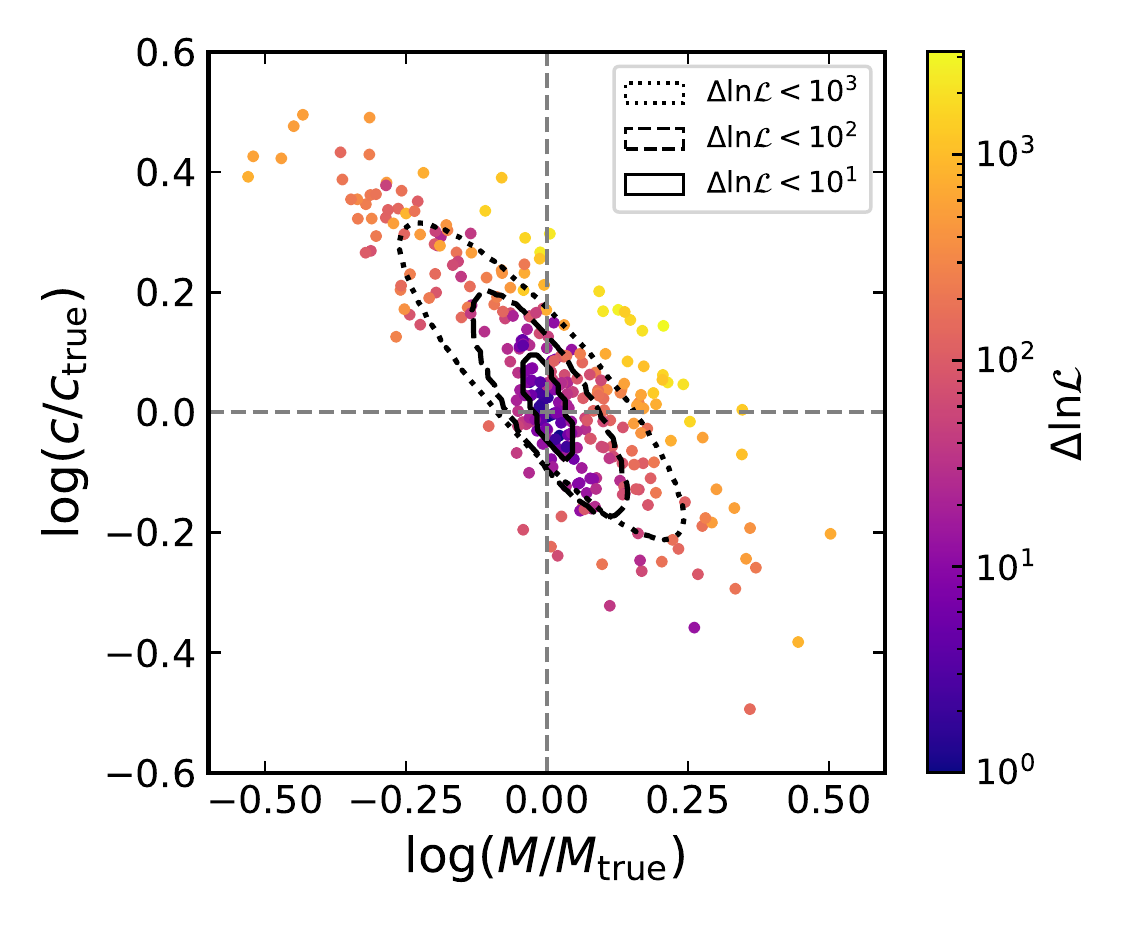}
    \caption{Best-fitting halo mass ($x$-axis) and concentration ($y$-axis) normalized by their true values for our sample of galaxy clusters. Each point shows the results for one cluster halo using the \oPDF estimator, and the points are colour coded by the log-likelihood difference, \deltaL, between the best-fitting and true parameters. The black solid, dashed and dotted ellipses represent the 1$\sigma$ dispersion of three sub-samples selected according to \deltaL\ $<10^1$, \deltaL\ $<10^2$ and \deltaL\ $<10^3$, respectively.}
    \label{fig:lnL}
\end{figure}

As shown in Fig.~\ref{fig:lnL}, \deltaL traces well the deviation of best-fitting from true parameters. The best-fitting mass and concentration parameters of each cluster halo in Fig.~\ref{fig:lnL} are estimated with \oPDF and using $10^5$ randomly selected dark matter particles between an inner radius of $200 h^{-1} \rm kpc$ and an out radius of $r_{200}$. Particles belonging to substructures are not used. We remove clusters whose best-fitting parameters lie outside the 3-$\sigma$ confidence region of the distribution in Fig.~\ref{fig:lnL} to eliminate catastrophic failures in the fits. This is done iteratively until the remaining number of clusters converge, resulting in 310 clusters in the end. The data points as colour coded by \deltaL show a smooth gradient over the parameter plane. In general, clusters with a low value of \deltaL have a small bias of best-fitting from true parameters. We further divide our sample of clusters into three subsamples according to \deltaL, \deltaL\ $<10^1$, \deltaL\ $<10^2$ and \deltaL\ $<10^3$, whose 1$\sigma$ dispersions are demonstrated by the black solid, dashed and dotted ellipses. The size of the 1$\sigma$ ellipse obviously increases with the increase in \deltaL. Moreover, there are two interesting patterns. First, there is a strong negative correlation between the fitted mass and concentration parameters. Besides, the distribution of the fits is not exactly symmetric along the direction orthogonal to the correlation. There are a few data points with \deltaL$>10^3$ on the top right corner, i.e., both halo mass and concentration are over-estimated, but not in the lower left corner. We suspect that the large bias along this direction may be caused by strong deviations from the steady-state assumption, and we will provide more detailed discussions in Section~\ref{sec:disc}. 

In addition to \deltaL, we have also tried a few other variables as targets for RFR, such as the normalized mean-phase deviation $\bar{\Theta}$~\citep{Han2016a}\footnote{Systems in steady states are expected to have uniform phase angle distributions.} and the ratio between best-fitting and true halo mass enclosed within the half radius of tracers. All these alternative choices of variables also demonstrate a clear correlation with the bias in the dynamical model (see Section~\ref{sec:disc} for details), but not as good as \deltaL, 
Thus we decide to choose \deltaL as the target for RFR.

\subsubsection{Features} 
\label{sec:features}

The \DS indicators or features we input to RFR can be divided into two different groups. The first group is constructed directly from our hydrodynamical simulations with full 3-dimensional (3D) information, while the other group is derived from the mock optical, X-ray and SZ maps in projection to mimic real observations.

Those features constructed directly from simulations in 3D include several traditional \DS indicators: 

\begin{enumerate}[label=\arabic*)]

\item Virial ratio ($\eta$)

$\eta\equiv(2T - E_{\mathrm{s}})/|W|$, where $T$ is the total kinetic energy, $E_{\rm s}$ is a surface pressure correction term and $W$ is the total potential energy \citep[applications in][for example]{Klypin2016,Cui2017}.

\item Centre-of-mass offset ($\Delta_{\rm r}$) 

$\Delta_{\mathrm{r}}\equiv |\vec{r}_{\rm CoM} - \vec{r}_{\rm den}|/r_{200}$, where $\vec{r}_{\rm CoM}$ is the coordinate of mass centre within $r_{200}$, and $\vec{r}_{\rm den}$ is the coordinate of the maximum density in the halo \citep[e.g.,][]{Duffy2008,Sembolini2014}.

\item Fraction of mass in subhalo ($f_{\rm s}$, see \citealp{Cui2018}).

\item Relaxation parameter ($\chi_{\rm DS}$).

A combination of $\eta$, $\Delta_{\rm r}$ and $f_{\rm s}$ \citep{Kuchner2020}:
\begin{equation}
    \chi_{\rm DS} = \sqrt{\frac{3}{\left(\frac{\Delta_{\rm r}}{0.04}\right)^2 + \left(\frac{f_{\rm s}}{0.1}\right)^2+ \left(\frac{|1-\eta|}{0.15}\right)^2}}, 
\end{equation}
where the critical values in the denominators (0.04, 0.1 and 0.15) are used to scale/unify different terms. 
More details can be found in \citet{Cui2018}, who adopted the cirteria of $0.85<\eta<1.15$, $\Delta_{\rm r}<0.04$ and $f_{\rm s}<0.1$ to select dynamically relaxed clusters.

\item Minor-to-major axis ratio of the inertial tensor ($c/a$), describing the shape of haloes \citep[e.g.,][\citetalias{paperI}]{Wang2017}.

\vspace{0.3cm}
We also include a few non-conventional features: 
\vspace{0.2cm}

\item The fraction of stellar mass in the brightest centre galaxy (BCG) plus ICL relative to total stellar mass ($f_{\rm ste}$)  

This is defined within $r_{500}$, i.e., the radius within which the mean matter density is 500 times the critical density of the universe. The stellar mass in ICL and BCGs are calculated using star particles not belonging to any satellite subhaloes. Despite the fact that the fraction of stellar mass in ICL as a proxy to the \DS is still under debates \citep[e.g.,][]{Pierini2008,Adami2013,Yolanda2018}, we still include it.

\item The age-weighted mass accretion rate ($\dot{M}_\mathrm{a}$):
\begin{equation}
    \dot{M}_\mathrm{a} = \frac{\sum_i t_i \Delta M/\Delta t}{\sum_i t_i},
\end{equation}
where $t_i$ is the age of universe at a given snapshot $i$, $\Delta M$ and $\Delta t$ are the difference in virial mass and the age of the universe between two neighbouring snapshots. 
$\dot{M}_\mathrm{a}$ is calculated using 15 snapshots between redshift $z=0$ and $z\sim0.36$. The age-weighted mass accretion rate has larger weights for more recent mass growth. \end{enumerate}

In addition to the above features which we focus on, we also try to include in our list a large number of extra features, in order to investigate what is the best that we can achieve. This not only helps to replenish the description of the \DS from different aspects, but also helps to test the robustness of RFR. These features include: 1-2) the current subhalo masses of the first and second most massive satellites, defined through all bound dark matter, star and gas particles; 3-4) the distances to halo centre\footnote{Unless otherwise specified, we define the halo centre as position corresponding to the matter density peak.} for the first and second most massive satellites; 5) the number fraction of red satellite galaxies\footnote{Red galaxies are defined as those objects whose $g-r$ colours are larger than 0.75.}; 6-7) the total mass in stars and gas component within $r_{200}$; 8) offset between the most bound particle and halo centre; 9) 3D total velocity dispersion; 10-12) spin parameters for all particles, gas and stars; 13-14) the metallicity in gas and stars; 15) the age of clusters; 16) the star formation rate (SFR) of all member galaxies in the cluster; 17) the mass-weighted temperature; 18) electron number density and 19) entropy of hot gas within $r_{500}$.     

In total, we have 26 potential \DS indicators constructed using full 3D information from the simulation.

The ICM morphological features constructed from the X-ray, SZ and optical maps are described in the following, and more details can be found in \citet{DeLuca2021}. These features are more closely linked to real observations. 
\begin{enumerate}[label=\arabic*)]

\item Asymmetry ($A$) 

This is the normalized difference between the original map and a rotated/flipped map \citep{Schade1995,Zhang2010}. For each cluster, we take the maximum difference with four different rotations/flips: $90^\circ$, $180^\circ$ rotations and horizontal and vertical flips.

\item Light concentration ratio ($c$), defined as the ratio between the flux enclosed within 40 kpc and 400 kpc\citep{Santos2008}.

\item Centroid shift ($w$). 

$w$ is defined as the average shifts in the centroids of different concentric circles with increasing apertures \citep{Mohr1993,Bohringer2010}.

\item Power ratio \citep[$P$;][]{Buote1995} 

This is the ratio between the third and zeroth order terms of the multipole decomposition of the ICM map.

\item Gaussian fit \citep[$G$;][]{Cialone2018}, which is defined as the ratio between the standard deviations ($\sigma_x$ and $\sigma_y$) of a 2D Gaussian fit to the X-ray and SZ maps.

\item Strip \citep[$S$;][]{Cialone2018}, which is defined as the normalized difference of light profiles with different angles passing through the centroid. We here consider four strips with angles equal to $0^\circ$, $45^\circ$, $90^\circ$ and $135^\circ$ following \citet{Cialone2018}.

\item Mixed feature $\mathcal{M}$

We also use a combination of all the six features above. Following \citep[][]{Cialone2018,DeLuca2021}, we adopt a weighted average of the six features, using the Kolmogorov-Smirnov(KS) p-values\footnote{The KS p-value is used to quantify the hypothesis that two samples are drawn from the same distribution. A larger p-value means the two samples are more likely to be drawn from different distributions.} as weights.
First, we calculate the distribution for each of the six features above, and for relaxed and unrelaxed clusters separately. Then for a given 
feature, the median p-values are evaluated for the distribution of relaxed and unrelaxed clusters with different apertures and at different redshifts ($0<z<1$). The minumum p-value selected from these median values is adopted as the weight. 
\end{enumerate}

Observationally, the centre of a galaxy cluster can be defined in a few different ways. The centre can be determined by the position of the BCG, which is defined as the intensity weighted centroid of the pixels associated with BCG. We identify BCGs within $0.5r_{500}$ in the $r$-band optical maps. Besides, the centres can be measured from the X-ray and SZ maps, which are usually defined as either the intensity weighted centroid position or the position of the pixel with the highest intensity (peak). These centroid positions defined in different ways can have offsets from each other, as well as from the maximum density location of the galaxy cluster in the simulation. We expect for unrelaxed galaxy clusters, the offsets are larger. In fact, these offsets are often considered as \DS indicators \citep[e.g.,][]{Rossetti2016,Lopes2018}. Hence we also include these position offsets as features. 

We denote the offset of the matter density peak from the BCG centre, from the intensity weighted center defined on SZ/X-ray maps and from the offset of intensity peak center on SZ/X-ray maps as $R_\mathrm{BCG}$, $R_\mathrm{SZ}$, $R_\mathrm{X}$, $R_\mathrm{Py}$ and $R_\mathrm{PX}$, respectively. The offsets for every two combinations of these centroid positions are denoted by $O_\mathrm{A-B}$, where $\mathrm{A-B}$ stands a particular combination. 

For convenience, we call the group of features constructed from simulations in 3-D as $D_3$, and the features from mock maps as $D_2$ throughout this paper. For the ICM morphological features in $D_2$, we denote them as $D_2$(X-ray or SZ, depending on the map used), while for those centroid position and offset related features, we denote them as $D_2$(Offset). We provide a summary of all features in Table.~\ref{tab:features}.

\begin{table*}
    \centering
    \caption{List of \DS indicators used in this work, constructed from 3D simulation data ($D_3$) and 2D mock maps ($D_2$). The indicators in mock maps contain three classes: those extracted from X-ray maps, those from SZ maps and those defined through offsets of peaks in optical, X-ray and SZ maps. See section~\ref{sec:features} for detail.}
    \label{tab:features}
    \begin{tabular}{c l c l c}
    \hline \hline
    \multicolumn{2}{c}{$D_3$}  & \multicolumn{3}{c}{$D_2$}\\
		\hline
      Feature & Description & Feature & Description & Class\\
      \hline
      $\eta$ & virial ratio & $A$ & asymmetry & X-ray/SZ\\
      $\Delta_{\rm r}$ & centre-of-mass offset & $c$ & light concentration ratio & X-ray/SZ\\
      $f_{\rm s}$ & fraction of mass in subhaloes & $P$ & power ratio & X-ray/SZ \\
      $\chi_{\rm DS}$ & combined relaxation parameter &  $w$ & centroid shift & X-ray/SZ\\
      $c/a$ & minor to major axis ratio of inertial tensor & $S$ & strip & X-ray/SZ\\
      $\dot{M}_\mathrm{a}$ & age-weighted mass accretion rate & $G$ & Gaussian fit & X-ray/SZ\\
      $f_{\rm ste}$ & stellar mass fraction of ICL and BCG within $r_{500}$ & $\mathcal{M}$ & combined morphological parameter & X-ray/SZ \\
      $M$1 & subhalo mass of the most massive satellite & $R_{\rm BCG}$ & BCG position & Offset \\
      $M$2 & subhalo mass of the second massive satellite & $R_{\rm SZ}$ & SZ centroid & Offset \\
      $r$1 & distance of the most massive satellite to centre & $R_{\rm X}$ & X-ray centroid & Offset \\
      $r$2 & distance of the second massive satellite to centre & $R_{\rm Py}$ & SZ peak & Offset \\
      $f_{\rm red}$ & number fraction of red galaxy & $R_{\rm PX}$ & X-ray peak & Offset \\
      $M_*$ & total stellar mass within $r_{200}$ & $O_{\rm BCG-y}$ & offset between BCG and SZ centroid & Offset \\
      $M_{\rm gas}$ & total gas mass within $r_{200}$ & $O_{\rm BCG-X}$ & offset between BCG and X-ray centroid & Offset \\
      $\Delta_{\rm mbp}$ & offset between the most bound particle and halo centre & $O_{\rm BCG-Py}$ & offset between BCG and SZ peak & Offset \\
      $\sigma_{\rm V}$ & 3D velocity dispersion & $O_{\rm BCG-PX}$ & offset between BCG and X-ray peak & Offset \\
      $\lambda_{\rm t}$ & spin parameter for all matters & $O_{\rm SZ-X}$ & offset between SZ centroid and X-ray centroid & Offset \\
      $\lambda_{\rm gas}$ & spin parameter for gas & $O_{\rm SZ-Py}$ & offset between SZ centroid and SZ peak & Offset \\
      $\lambda_{\rm star}$ & spin parameter for star & $O_{\rm SZ-PX}$ & offset between SZ centroid and X-ray peak & Offset \\
      $Z_{\rm gas}$ & gas metallicity & $O_{\rm X-Py}$ & offset between X-ray centroid and SZ peak & Offset \\
      $Z_{\rm star}$ & star metallicity & $O_{\rm X-PX}$ & offset between X-ray centroid and X-ray peak & Offset \\
      $\rm SFR$ & cluster SFR & $O_{\rm Py-PX}$ & offset between SZ peak and X-ray peak & Offset \\
      $\rm age$ & cluster age \\
      $T_{500}$ & mass-weighted temperature within $r_{500}$\\
      $n_{\rm e}$ & electron number density within $r_{500}$\\
      $K_{500}$ & entropy within $r_{500}$\\ 
    \hline
\end{tabular}
\end{table*}

\subsubsection{OOB score}

When constructing a forest, each tree is trained against a bootstrap realization of the parent training sample. This means about 36.8\% of data in the training sample are not used to construct the tree in the large sample limit. These unused data are called the \OOB data. Thus, the \OOB data can be used to evaluate quality of the learning outcome and the correlation between the target variable and the input features. Based on the \OOB data, we evaluate the RFR performance with a popular and important metric, the coefficient of determination, $R^2$, which is defined as:  
\begin{equation}
    R^2 = 1 - \frac{\sum_i \left(y_i - \hat{y}_i \right)^2}{\sum_i \left(y_i - \bar{y}_i \right)^2},
\end{equation}
where $y_i$ is the target variable, $\hat{y}_i$ is the prediction from \RFR and $\bar{y}_i$ is the average value in the data. The $R^2$ calculated from the \OOB data is referred to as the \OOB score. The \OOB score of different trees in the forest will be averaged, leading to the final \OOB score of the learning outcome. $R^2$ can be interpreted as the fraction of scatter in $y$ that can be explained by the model. A close to unity value of \OOB score means that the features are strongly correlated with the target, and the scatter between predicted and true target values are well controlled. 

In our analysis, the importance of each individual feature is determined by the \OOB score when the target variable is trained and predicted using only this particular feature. Note RFR can also output feature importance ranking for a group of features, when these features are jointly used as inputs to predict the target variable. However, the default feature importance ranking may suffer from a so-called masking effect, when there are strong correlations amount these features  \citep[see e.g.][]{Shi2021}. Thus we choose not to use the default importance ranking returned by RFR.

\section{Results} \label{sec:performance}

In this section, we investigate the importance of different features in determining the \DS of galaxy clusters. As we have mentioned, we represent the intrinsic \DS of a galaxy cluster with the log-likelihood difference between the best-fitting and true halo parameters, \deltaL, and the importance of each feature is quantified by the \OOB score when this particular feature is used to train and predict \deltaL with RFR. First, we show the overall results using all $D_3$ and $D_2$ features. Then, we will investigate the importance of each individual feature as a proxy to the \DS and the correlations among different features. Moreover, the importance of different features can tell what is the most dominant source of systematics in dynamical modelling (e.g. violation of steady-state or spherical assumptions). Finally, we explore the best feature combinations in predicting the \DS.

\subsection{Basic performance} \label{sec:basic}
\begin{figure}
    \centering
    \includegraphics[width=0.45\textwidth]{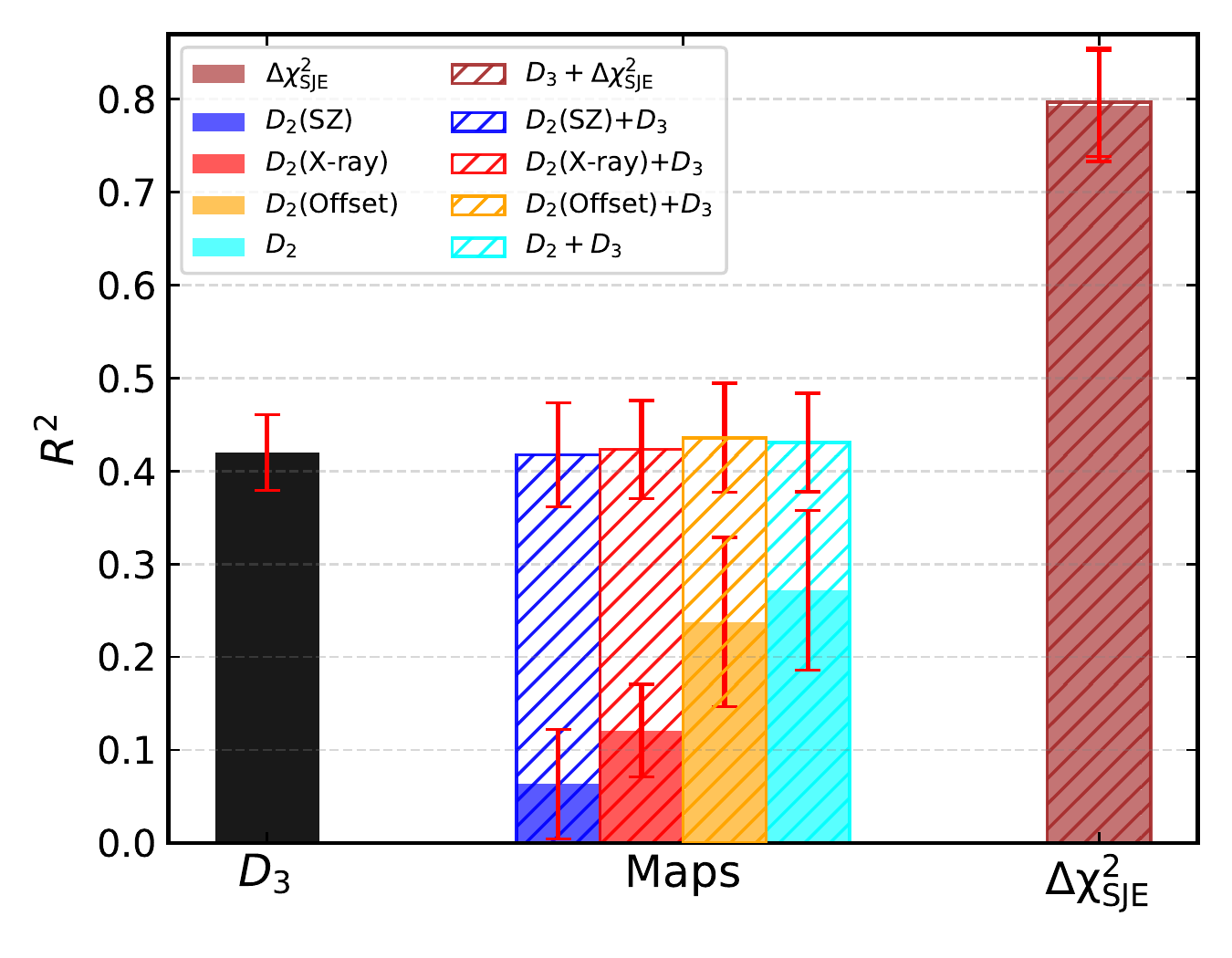}
    \caption{The \OOB score of the learning outcome based on features in different categories. The black column on the left shows the \OOB score when all features in the $D_3$ category are used. The colour columns in the middle are based on different groups of $D_2$ features (filled) or these $D_2$ features combined with all $D_3$ features (dashed). Blue, red, orange and cyan columns refer to results based on $D_2$ features measured from the SZ map, $D_2$ features from the X-ray map, offsets between centroid positions defined in different ways in the $D_2$ category and all $D_2$ features, respectively. In addition to the default dynamical modelling method used in this paper (oPDF), we have also tried the SJE modelling. The brown column on the right is the \OOB score for the learning outcome when the $\chi^2$ difference between the best-fitting and true SJEs is used as the input feature. The \OOB scores are averaged over 50 randomly selected training samples, while the errorbars show the estimated uncertainties of the scores (see text for detail).  
    }
    \label{fig:total}
\end{figure}

In Fig.~\ref{fig:total}, we show the \OOB score in the learning outcome when we input all $D_3$ features, different categories of $D_2$ features and different combinations of $D_2$ and $D_3$ features for training. The training sample is randomly selected for 50 times, and the \OOB scores we report in this paper are the average values over the 50 samples. Note that these 50 samples are not independent, but resemble a jackknife resampling with a 30\% hold-out each time. As a result, the uncertainty of the resulting \OOB score is estimated as the scatter among the 50 samples multiplied by a factor of $\sqrt{3-1}\sim1.44$ following the jackknife approach. We have verified using Monte-Carlo samples that the uncertainty calculated in this way is a fair estimate of the true error. 

When all features in the $D_3$ category are used, the \OOB score is about 0.42 (black filled column on the left). For filled columns with different colours in the middle, it seems features measured from SZ maps are slightly less important than those measured from X-ray maps (blue and red filled columns in the middle respectively), but both have very low \OOB scores ($<\sim 0.12$). $D_2$ features defined through the offsets between different types of centroid positions are more important, with an \OOB score of $\sim0.24$ (orange filled column), indicating they are better proxies to the \DS of galaxy clusters than the geometrical features measured from SZ or X-ray maps. However, even after including all available $D_2$ features, the \OOB score is only 0.27 (cyan filled column), which is significantly smaller than the score when all features in the $D_3$ category are used. 
This implies that features measured from observed optical, SZ and X-ray maps in projection carry less information about the \DS of galaxy clusters than those features directly calculated from the simulation in 3-dimension.

Combining $D_3$ and $D_2$ features, the \OOB scores as shown by the dashed colour columns in the middle are all significantly increased, reaching similar levels as the case when all $D_3$ features are used. This implies that $D_2$ features only cover part of the information about the \DS of galaxy clusters, and the information is already almost fully contained in $D_3$ features. Note our mock maps are free from sky contamination or instrumental noise. The loss of information in the $D_2$ set compared with $D_3$ may be due to the limited amount of information carried in the SZ, X-ray or optical channels of the baryonic components, or due to the projection from 3D to 2D.  

However, the \OOB score is still low even after using all available features in $D_3$ and $D_2$ categories. Perhaps there are other important features, which are not included in our analysis but contain a non-negligible fraction of information about the \DS. 
To check what is the best one can achieve, we construct another feature, $\Delta \chi_{\rm SJE}^2$, to explore the missing part of information. $\Delta \chi_{\rm SJE}^2$ is the $\chi^2$ variable between the best-fitting and true halo parameters based on the Spherical Jeans Equation (SJE). After applying the SJE to each galaxy cluster in our analysis, $\Delta \chi_{\rm SJE}^2$ can be estimated for each of them, in a similar manner as how we estimate the log-likelihood ratio, \deltaL, for the \oPDF method (see details in \citetalias{paperI}). Since SJE and \oPDF adopt the same steady-state and spherical assumptions, we expect $\Delta \chi_{\rm SJE}^2$ to carry the full amount of information about the \DS as having been contained in \deltaL. After using $\Delta \chi_{\rm SJE}^2$ as the only input feature to learn and predict the target variable of \deltaL, we find that the \OOB score significantly increases to $\sim$0.8. If using $\chi_{\rm SJE}^2$ together with all features in the $D_3$ category, the \OOB score still remains $\sim$0.8. 

There is about 0.2 of difference before reaching a full score of unity. This can be attributed to the random nature of the target variable. As we have explained before, the target variable, \deltaL, is itself a random variable with statistical fluctuations. Such fluctuations are due to the finite number of tracers used in the dynamical model so that any statistical inference comes with a finite uncertainty. This statistical fluctuation is modulated by the amount of systematics in the halo, so that larger systematics correspond to a larger \deltaL \emph{on average}. With this understanding, we expect any physical model can at most predict the average \deltaL subject to its statistical fluctuations. 

\begin{figure}
    \centering
    \includegraphics[width=0.45\textwidth]{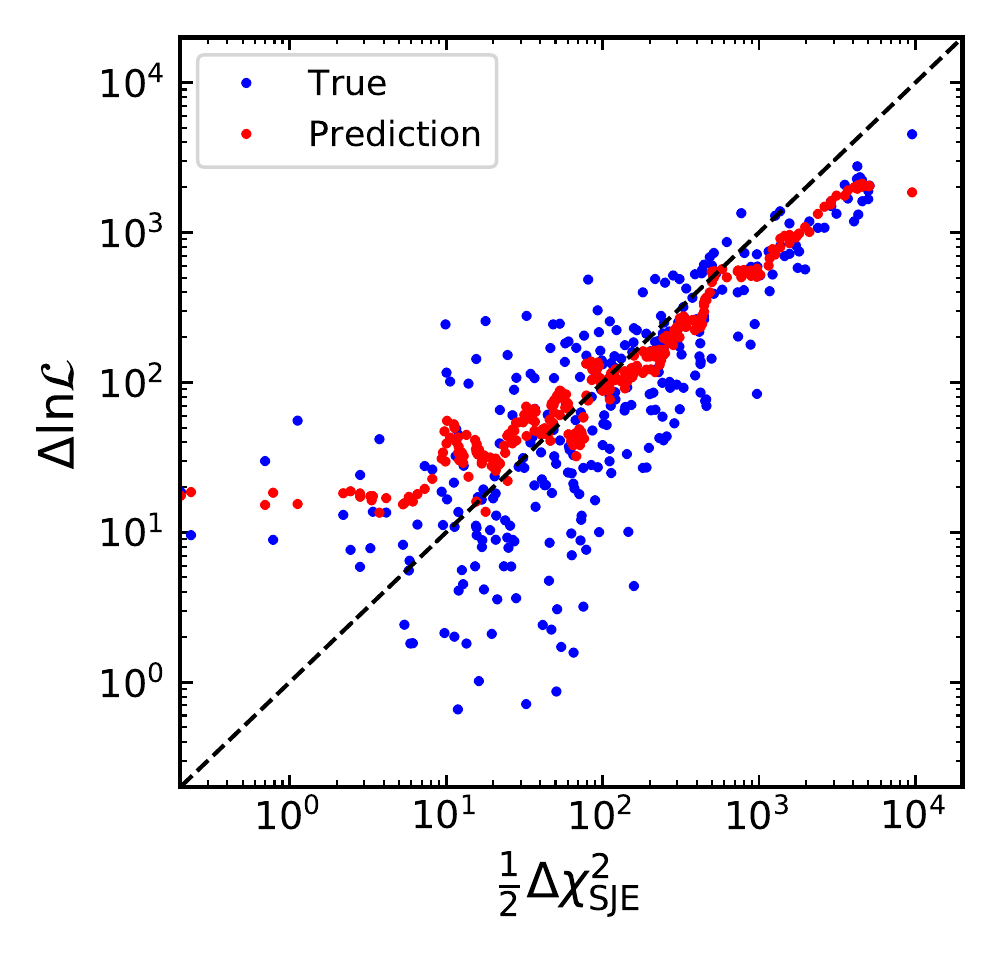}
    \caption{The log-likelihood ratio, \deltaL, between the best-fitting and true halo parameters of the \oPDF method, versus half of the $\chi^2$ quantity between best-fitting and true halo parameters ($\Delta \chi_{\rm SJE}^2$) of the SJE modelling. Each data point represents one galaxy cluster. The true values of \deltaL and predicted \deltaL from RFR using only $\Delta \chi_{\rm SJE}^2$ as the feature are shown in blue and red colours, respectively.}
    \label{fig:error}
\end{figure}

Fig.~\ref{fig:error} demonstrates this through the relation between \deltaL and $\chi_{\rm SJE}^2$. The true values of \deltaL and $\frac{1}{2}\Delta \chi_{\rm SJE}^2$ show a strong linear relation, but also with large scatters. The linear trend reflects the intrinsic part, while the scatter is a representation of the random errors. Meanwhile, we show the prediction of \deltaL (red points) given by RFR only using $\Delta \chi_{\rm SJE}^2$ as the input feature. The linear relation between the predicated $\Delta \ln \mathcal{\hat{L}}$ and $\frac{1}{2}\Delta \chi_{\rm SJE}^2$ is obvious but the scatter becomes negligible. This indicates that the intrinsic ratio is recognised, while the random errors are not captured by the learning process of RFR. To conclude, due to the fact that the log-likelihood ratio or the $\chi^2$ quantity is a random variable, there exists a upper limit to the \OOB score of our model even after using all available features. 

\subsection{Importances of individual features}
\label{sec:imp_individual}

As illustrated in Section.~\ref{sec:features}, a few features in the $D_3$ category might have very weak correlations with the actual \DS of galaxy clusters, and the inclusion of these features would introduce noise during the RFR learning process. Besides, in previous studies, a single feature is often adopted as the proxy to the \DS of galaxy clusters, both observationally and in numerical simulations, and thus it is very important to understand which feature is better than the others. In this section, we explore the importances of individual features in determining the \DS of galaxy clusters.

\begin{figure*}
    \centering
    \includegraphics[width=\textwidth]{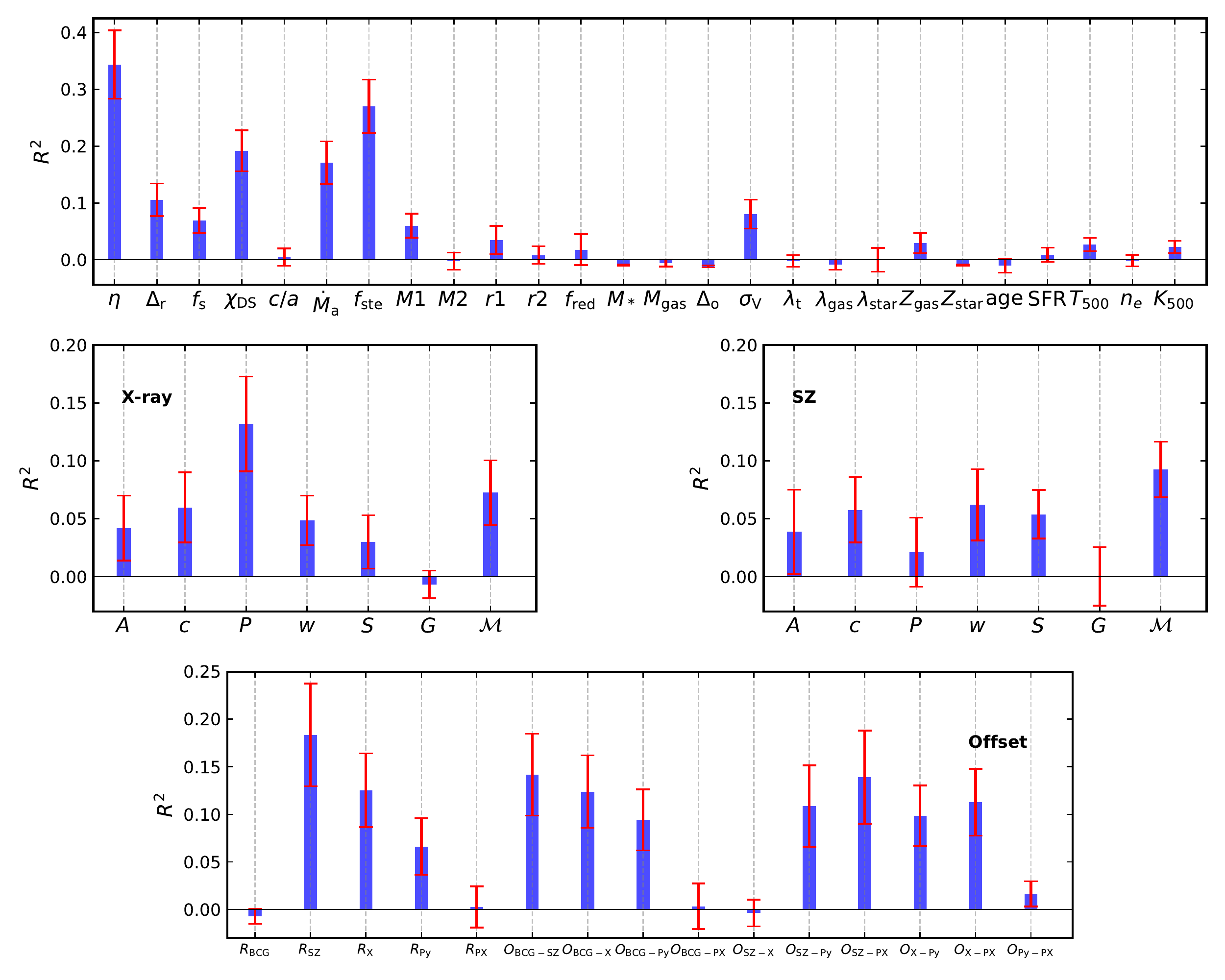}
    \caption{Blue histograms show the \OOB scores of different features, with red errorbars showing their uncertainties. Note a negative \OOB score means a very low feature importance, i.e., the prediction of the target variable based on this feature is even worse than simply using the mean for the prediction. The top panel shows results for features in the $D_3$ category. Middle panels show results for features measured from the SZ and X-ray maps in the $D_2$ category. The bottom panel is for all centroid position and position offsets related features. }
    \label{fig:one}
\end{figure*}

In the top panel of Fig.~\ref{fig:one}, we show the \OOB score for each individual feature in the $D_3$ category. Among all the features, the virial ratio, $\eta$, presents the highest \OOB score ($\sim$ 0.34). Besides, the fraction of stellar mass in the BCG and ICL, $f_{\rm ste}$, the combined relaxation parameter, $\chi_\mathrm{DS}$, the age-weighted mass accretion rate, $\dot{M}_\mathrm{a}$, the centre-of-mass offset, $\Delta_{\rm r}$, the 3D velocity dispersion, $\sigma_{\rm V}$, and the fraction of mass in subhaloes, $f_{\rm s}$ have relatively high \OOB scores. 

\begin{figure}
    \centering
    \includegraphics[width=0.5\textwidth]{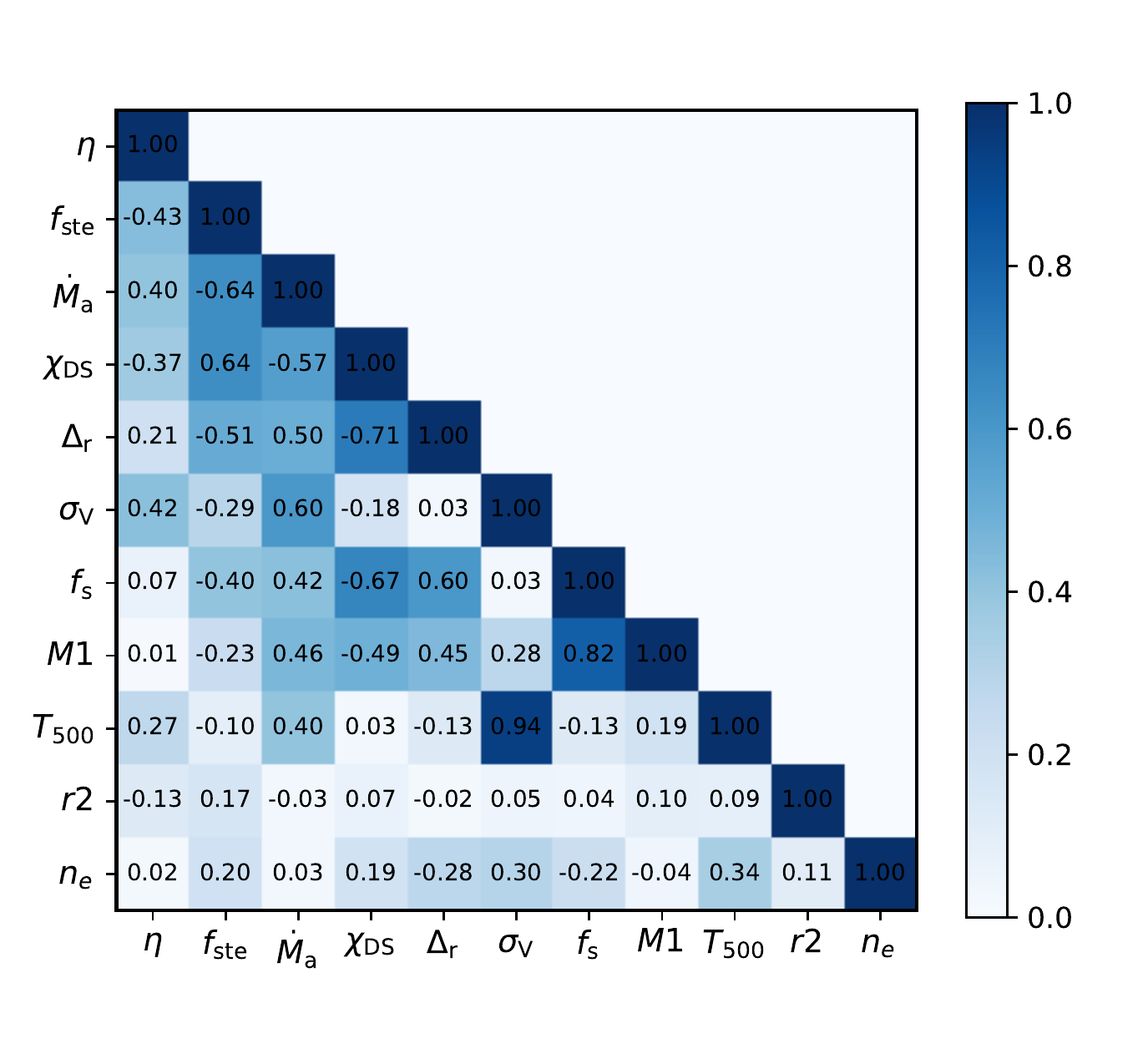}
    \caption{The Pearson correlation coefficients among eleven features selected from the $D_3$ category. The colour indicates the amount of correlation, with the exact value of the correlation coefficient marked in each grid. Features are ordered in the axis according to decreasing \OOB scores from left to right and top to bottom.}
    \label{fig:cc1}
\end{figure}

Naturally, we expect the current \DS of a galaxy cluster to have a strong dependence on its mass accretion history, i.e., clusters which assembled earlier are on average more relaxed today. Thus the importance of $\dot{M}_\mathrm{a}$ is high. 
In Figure~\ref{fig:cc1}, we show the Pearson correlation coefficients of a few representative features in the $D_3$ category. The features listed from left to right in the $x$-axis are ordered by their individual \OOB score. The Pearson correlation coefficient is a measure of linear correlation between two data sets and quantifies the steepness of one feature as a function of another feature after normalisation \citep{Han2019}. $f_{\rm ste}$, $\sigma_{\rm V}$, $\chi_\mathrm{DS}$ and $\Delta_{\rm r}$ all show strong positive or negative correlations with each other and also with $\dot{M}_\mathrm{a}$, indicating they all carry a rich amount of information about the mass accretion history. The correlation coefficient between $f_{\rm s}$ and $\dot{M}_\mathrm{a}$ is positive but smaller than 0.5, but $f_{\rm s}$ has stronger correlations with both $\chi_\mathrm{DS}$ and $\Delta_{\rm r}$. The correlation between $\eta$ and a few other features is relatively weak in Figure~\ref{fig:cc1}, implying $\eta$ carries more {\it independent} information on the \DS than the other features. 

However, some physical features such as the SFR, age, stellar mass, gas mass, metallicity, the spin parameters of the stellar and gas components, the mass-weighted temperature, the electron number density, the virial mass and distance of the second most massive satellite all seem to be relatively poor indicators of the \DS, with low or even negative \OOB scores, despite the fact that part of these features are observable. 

Interestingly, the minor to major axis ratio, $c/a$, shows a quite low \OOB score. Most dynamical models adopt steady-state and spherical assumptions, which may cause significant systematic errors if a system deviates from the assumptions. It might be more severe for cluster haloes, because they are formed later than galactic haloes. Our results here show that \deltaL depends very weakly on $c/a$ for cluster haloes, which quantifies the shape of the host halo. Similar conclusions have been reached by \cite{Wang2015} and \cite{Rehemtulla2022} for galactic haloes and based on different modelling approaches. On the other hand, \deltaL shows much stronger dependence on $\eta$, $\chi_{\rm DS}$ and $\Delta_{\rm r}$, which are good proxies to the \DS of the system, reflecting the steady state. Hence the deviation from spherical symmetry is less important compared with the deviation from the steady state. As has been shown in \citetalias{paperI}, the difference between the best-fitting and true halo parameters for two subsamples divided by $c/a$ is small, whereas the difference is more significant for two subsamples split according to the relaxation criteria\footnote{\citet{Cui2018} treat galaxy clusters with $0.85 < \eta < 1.15$, $\Delta_{\rm r} < 0.04$ and $f_{\rm s} < 0.1$ as relaxed systems.} of \citet{Cui2018}. These are in  very good agreements with what we have found with the RFR approach.

We also investigate the ICM morphological features in the three subsets of the $D_2$ category. The \OOB scores of the same features can significantly vary between X-ray and SZ maps. The power ratio, $P$, measured from X-ray maps, shows the highest \OOB score, but the score is significantly decreased when this feature is measured from SZ maps. The combined morphological parameter, $\mathcal{M}$, preserves a high level of importance in both X-ray and SZ maps, which is also investigated and discussed in \citet{DeLuca2021}. For position and position offset features, the intensity weighted centroid position of the BCG based on the optical maps, $R_\mathrm{BCG}$, has a negative \OOB score, whereas the intensity weighted centroid positions based on the X-ray and SZ maps ($R_{\rm X}$ and $R_{\rm SZ}$) are significantly more important. $R_{\rm SZ}$ presents the highest \OOB score, and the position offsets involving either $R_{\rm X}$ or $R_{\rm SZ}$ show high \OOB scores as well. On the contrary, the importance of the centroid positions based on the peak intensity of X-ray and SZ maps ($R_{\rm PX}$ and $R_{\rm Py}$) are significantly lower. Our findings are consistent with the results of \citet{DeLuca2021}.

Based on the analysis above, we conclude that the virial ratio, $\eta$, or the age-weighted mass accretion ratio, $\dot{M}_\mathrm{a}$, are good proxies to indicate the \DS of galaxy clusters in numerical simulations. Observationally, it is helpful to use the offsets between different types of centroid positions to indicate the \DS, especially the offsets involving the usage of the intensity weighted centroid positions measured from X-ray and SZ maps. However, as we will move on to talk about the importance of different combinations of features, we will show that it could be risky to use only one feature to quantify the \DS.

\subsection{The best feature combinations}
\label{sec:imp_comb}

As shown in Fig.~\ref{fig:one}, the \OOB score of any individual feature does not reach the maximal value when all features are included (see Fig.\ref{fig:total}), indicating an incomplete description of the \DS based on single features. However, using a large number of features to indicate the \DS is inconvenient in practice, and thus a combination of a few features is the most reliable and practical choice. Based on this idea, we investigate how well the \DS can be described with combinations of different numbers of features.

\begin{figure}
    \centering
    \includegraphics[width=0.45\textwidth]{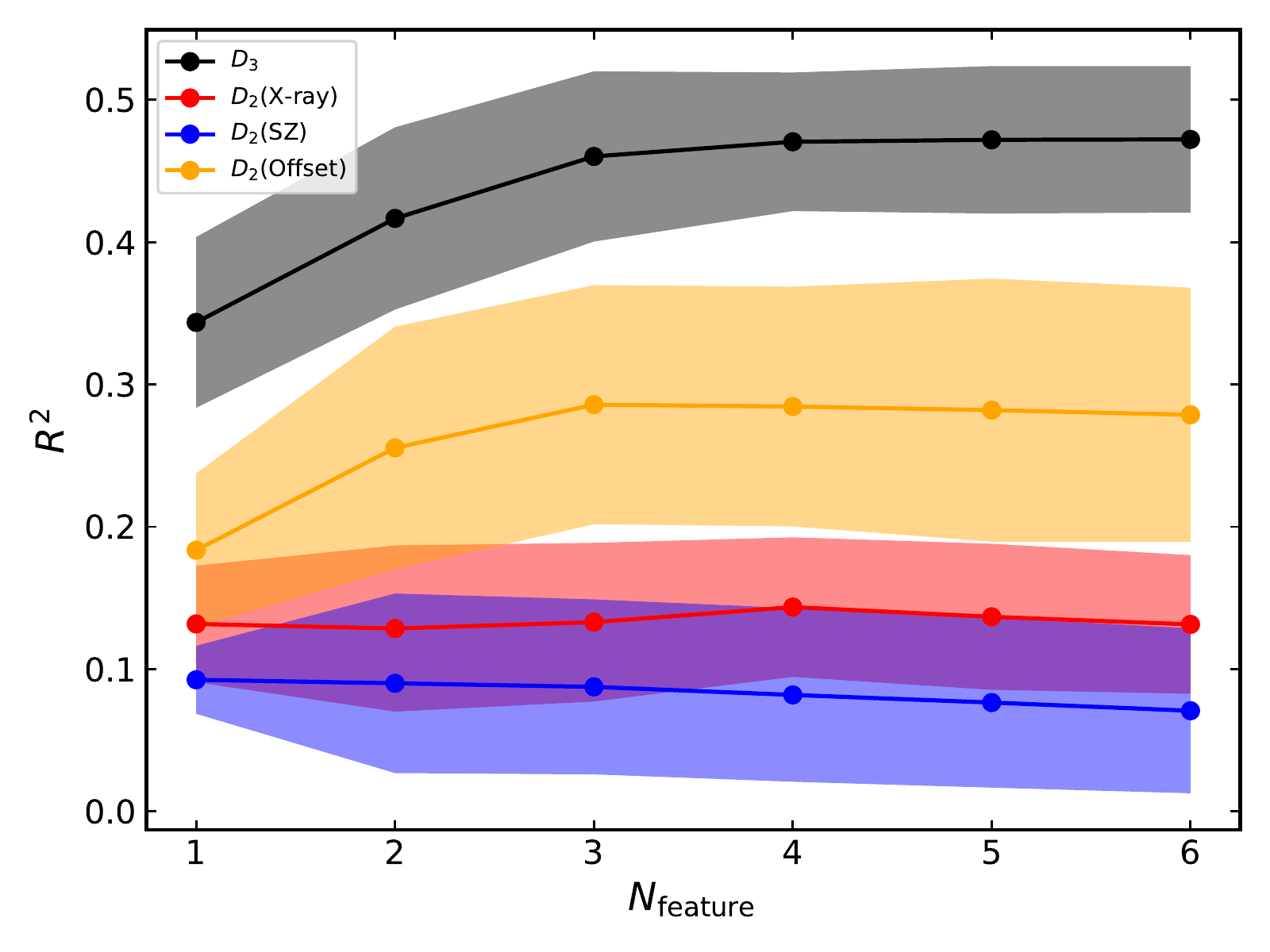}
    \caption{The highest \OOB score in all combinations of a different number of features. The black, red, blue and orange points respectively show the mean \OOB scores using the $D_3$, $D_2$(X-ray), $D_2$(SZ) and $D_2$(Offset) feature categories. The shadow regions represent the $1\sigma$ error of \OOB score generated with 50 randomly selected training samples.}
    \label{fig:all}
\end{figure}

In Fig.~\ref{fig:all}, we present the highest \OOB scores that can be reached and with all possible combinations of a certain number of features. The shaded regions represent the $1\sigma$ uncertainties of 50 randomly selected training samples. 
The \OOB scores for features in the $D_3$ category are higher than the other three subsets of $D_2$ category and at any number of features. The combination of features defined through position and position offsets in the $D_2$ category presents higher scores compared with the combination of features measured from X-ray or SZ maps. 

For combinations of features in the $D_3$ category and features defined through position and position offsets in the $D_2$ category, the \OOB scores first increase with the number of included features, but quickly converge at about three features. A combination of four or five features show very similar performances compared to the usage of only three features. We note that the highest \OOB score when six features in the $D_3$ category are used is $\sim$ 0.47, which is slightly higher than the score when all $D_3$ features are used (Fig.~\ref{fig:total}). The difference is because that the higher purity of features avoids the noise from some useless features. 

For features measured from X-ray and SZ maps, the \OOB scores stay nearly constant, despite the increase in the number of features. This means the top feature in either category ($P$ in X-ray and $\mathcal{M}$ in SZ) has already contained almost all the information about the \DS that may be carried in other features of the same category. 

In Table~\ref{tab:score}, we list the top five most important feature combinations for single, double and triple features  from the $D_3$ category. Consistent with Section~\ref{sec:imp_individual}, the virial ratio, $\eta$, is the most important single feature, and it appears in almost all the different combinations. For the ranking of double features, the top four most important combinations are $\eta$ together with the second, third, fourth and fifth most important individual features. However, the fifth most important two feature combination is $\eta$ plus the temperature, $T_{500}$. As an individual feature, $T_{500}$ has a lower importance than $\sigma_{\rm V}$, $f_{\rm s}$ and $M1$ particularly (see Figure~\ref{fig:one}), but when it is combined with $\eta$, the importance ranking is higher than $\eta$ combined with the other features. This is perhaps because $T_{500}$ contains more independent information, while $\sigma_{\rm V}$, $f_{\rm s}$ and $M1$ are more strongly correlated with $\eta$. According to Figure~\ref{fig:cc1}, this is true for $\sigma_{\rm V}$, but not exactly true for $f_{\rm s}$ and $M1$, which have smaller Pearson correlation coefficients with $\eta$ than that between $T_{500}$ and $\eta$. Maybe this is because the Pearson correlation coefficient quantifies the linear correlation part, while the non-linear correlations are not fully captured. In addition, we note the significance of the feature importance ranking is low if considering the typical 1-$\sigma$ uncertainties in Table~\ref{tab:score}. 

When the number of features increases to three, the distance to halo center for the second most massive satellite, $r2$, the electron number density, $n_{\rm e}$, and the 3-D velocity dispersion, $\sigma_{\rm V}$, come into the top five most important combinations, in addition to $T_{500}$ discussed above. Indeed, as shown by Figure~\ref{fig:cc1}, $r2$ and $n_{\rm e}$ tend to show weaker correlations with $\eta$ or $f_{\rm ste}$ than the correlations [$\chi_{\rm DS}$, $\eta$], [$\chi_{\rm DS}$, $f_{\rm ste}$], [$\dot{M}_\mathrm{a}$, $\eta$] and [$\dot{M}_\mathrm{a}$, $f_{\rm ste}$]. $\eta$ and $f_{\rm ste}$ appear in the first, second and fifth rankings of the three feature combinations, while $\eta$ appears in the third ranking, but not in the fourth ranking. In the fourth ranking, $\chi_{\rm DS}$ and $r2$ perhaps carry more independent information with each other, and thus the combination of $\chi_{\rm DS}$ and $r2$ is slightly more important than the combination of $\eta$ and $f_{\rm ste}$ or $\eta$ alone. Indeed, the correlation coefficient between $\chi_{\rm DS}$ and $r2$ is smaller than that between $\eta$ and $f_{\rm ste}$. $\sigma_{\rm V}$ appears in the fifth ranking, which shows slightly stronger correlation with $\eta$ or $f_{\rm ste}$ than the correlation [$T_{500}$, $\eta$], [$T_{500}$, $f_{\rm ste}$], [$n_{\rm e}$, $\eta$] and [$n_{\rm e}$, $f_{\rm ste}$], hence carrying less independent information than $T_{500}$ and $n_{\rm e}$. We note again that the rankings are not very significant compared with the typical 1-$\sigma$ uncertainties.

Based on our discussions above and the Pearson correlation coefficients, we group our features into four general classes: i) features quantifying the current dynamical status (e.g., $\eta$ and $\chi_{\rm DS}$); ii) features quantifying the merger history (e.g., $\dot{M}_\mathrm{a}$ and $f_{\rm ste}$\footnote{We group $f_{\rm ste}$ together with $\dot{M}_\mathrm{a}$, because $f_{\rm ste}$ has a strong correlation with $\dot{M}_\mathrm{a}$ and $1-f_{\rm ste}$ is the fraction of stellar mass locked in surviving satellites, which is closely connected to the halo assembly history.}); iii) the properties of satellites/subhaloes (e.g., $r2$); and iv) the properties of ICM (e.g., $T_{500} $ and $n_{\rm e}$). The \DS of galaxy clusters can be well captured by using a combination of three different features provided in Table~\ref{tab:score}, which is usually a combination of at least three different types of features above, with features containing the information of the present dynamical status of galaxy clusters and halo evolution histories usually having a high priority. 

\begin{figure}
    \centering
    \includegraphics[width=0.5\textwidth]{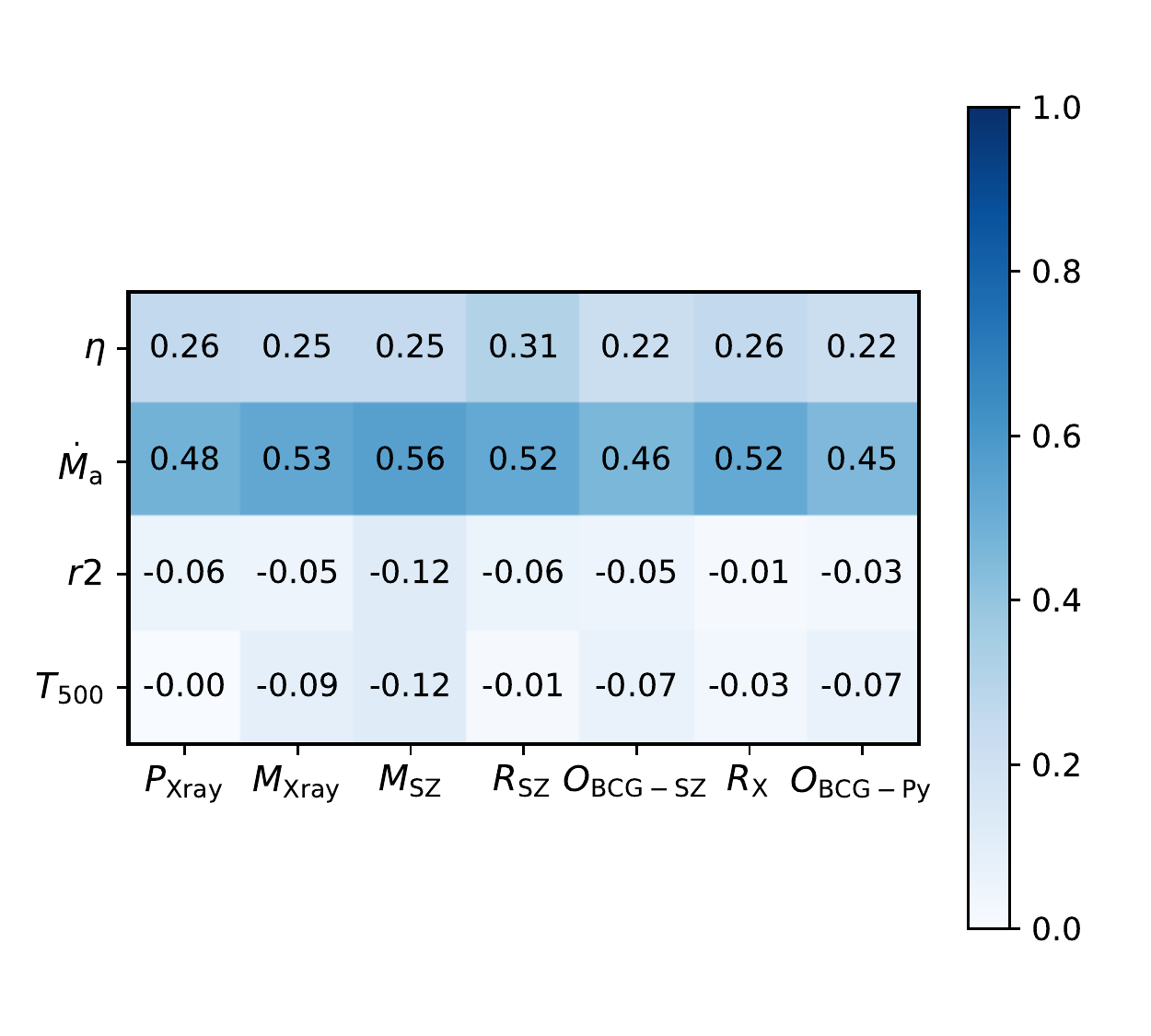}
    \caption{The Pearson correlation coefficients between four representative features from the $D_3$ category and seven projected features selected from the $D_2$ category. The colour indicates the amount of correlation, with the exact value of the correlation coefficient marked in each grid}
    \label{fig:cc2}
\end{figure}

Lastly, we select seven important morphological features ($P_{\rm Xray}$, $M_{\rm Xray}$, $M_{\rm SZ}$, $R_{\rm SZ}$, $O_{\rm BCG-SZ}$, $R_{\rm X}$ and $O_{\rm BCG-Py}$) based on their individual performance in Figure~\ref{fig:one}, and calculate their Pearson correlation coefficients with four representative features from the four classes defined above ($\eta$, $\dot{M}_\mathrm{a}$, $r2$ and $T_{500}$) shown in Fig.~\ref{fig:cc2}. Interestingly, all the seven morphological features show relatively stronger correlations with $\dot{M}_\mathrm{a}$. 

\begin{table*}
    \centering
    \caption{The top five highest \OOB scores in all combinations of different number of $D_3$ feature category. The list of the features is labelled at the top of the scores. The \OOB scores are listed as the average value over 50 randomly selected training samples, and the errors quoted here are the standard deviations among 50 samples multiplied by $\sqrt{2}$ (see the beginning of section~\ref{sec:basic} for details).}
    \label{tab:score}
    \begin{tabular}{c c c c c c}
    \hline \hline
      $N_{\rm feature}$& First & Second & Third & Fourth & Fifth\\
     \hline
      \multirow{2}{*}{ 1 } & $\eta$  & $f_{\rm ste}$ & $\chi_{\rm DS}$ & $ 
      \dot{M}_\mathrm{a}$ & $\Delta_{\rm r}$ \\
      &  $0.344$$\pm$0.060 & $0.270$$\pm$0.047 & $0.192$$\pm$0.036 & $0.171$$\pm$0.038 & $0.105$$\pm$0.029 \\
     \hline
     
      \multirow{2}{*}{2} & $\eta,f_{\rm ste}$ &  $\eta,\chi_{\rm DS}$ & $\eta, \dot{M}_\mathrm{a}$ & $\eta, \Delta_{\rm r}$  &  $\eta,T_{500}$\\
      &  0.417$\pm$0.064 & 0.380$\pm$0.064 & 0.377$\pm$0.069 & 0.367$\pm$0.058 & 0.363$\pm$0.078 \\
     \hline
     
      \multirow{2}{*}{ 3 } & $\eta,f_{\rm ste},T_{500}$  & $\eta,f_{\rm ste},n_{\rm e}$ & $\eta,\Delta_{\rm r},T_{500}$ & $\chi_{\rm DS},r2,T_{500}$  & $\eta,f_{\rm ste},\sigma_{\rm V}$  \\
      &  0.460$\pm$0.059 & 0.441$\pm$0.055 & 0.438$\pm$0.057 & 0.436$\pm$0.062 & 0.431$\pm$0.062 \\
    \hline
    \end{tabular}
\end{table*}

\section{Discussions} \label{sec:disc}

Based on the multivariate analysis of dynamical indicators, we have explored the importance rankings of individual features and of different feature combinations. We have shown that the virial ratio, $\eta$, is the most important feature which carries the largest amount of information on the \DS of galaxy clusters. Hence in this section we particularly select $\eta$ as a proxy and discuss how $\eta$ affects the bias in dynamical modelling. Besides, it was well recognised that the mass within the median radius of tracers, $M(r_{1/2})$, can be better constrained through dynamical modelling than the mass at other radii \citep[e.g.][]{Wolf2010,Walker2011,Han2016a}. In this section, we will explicitly show the correlation between $\eta$ and the bias in $M(r_{1/2})$. We will demonstrate that for systems with significantly large or small values of $\eta$, which are strongly out of equilibrium, the bias in $M(r_{1/2})$ is significant as well. 


\begin{figure*}
    \centering
    \includegraphics[width=\textwidth]{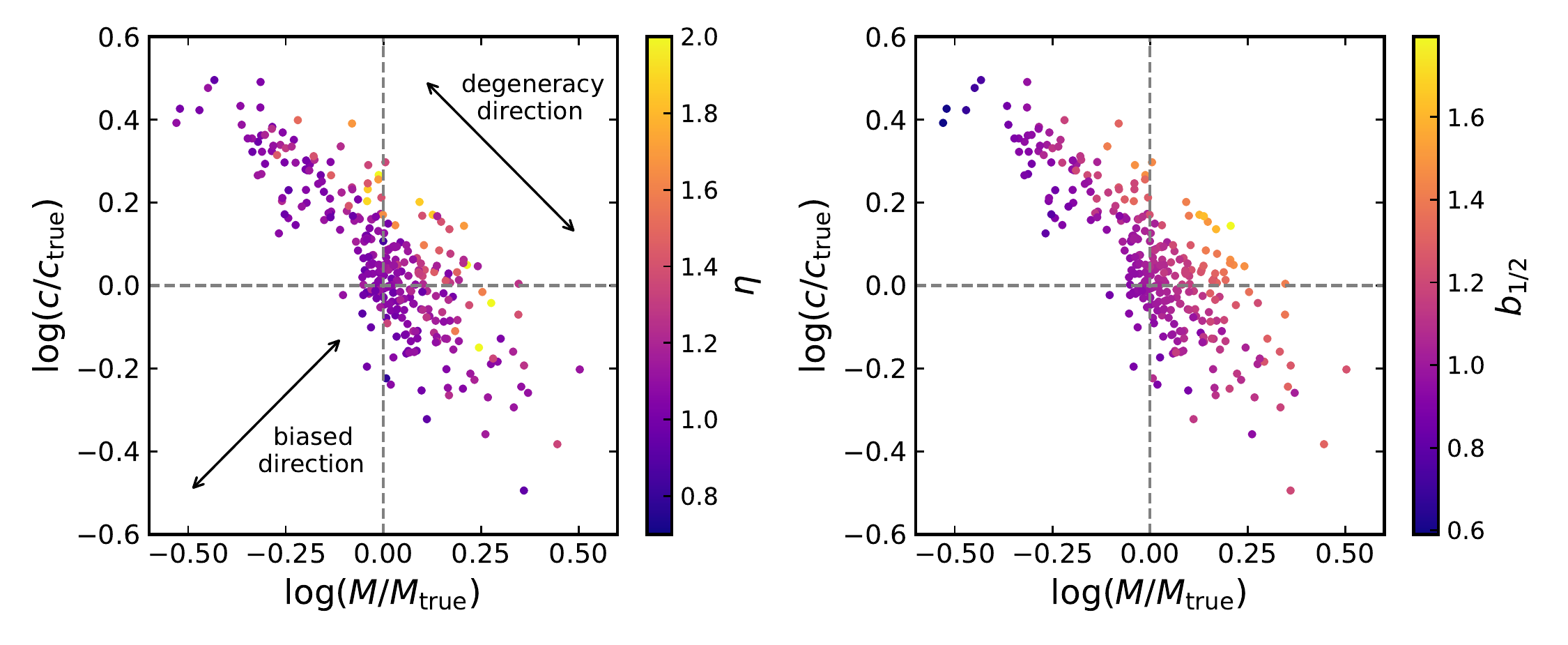}
    \caption{Best-fitting versus true halo mass ($x$-axis) and concentration ($y$-axis), colour coded by the virial ratio (left), $\eta$, and the half-mass bias (right), $b_{\rm 1/2}$. The arrows in the upper right and lower left corners respectively indicate the degeneracy and biased directions we defined.}
    \label{fig:2fea}
\end{figure*}


In the left panel of Fig.~\ref{fig:2fea}, we show the best-fitting halo mass ($x$-axis) and concentration ($y$-axis) normalized by their true values. Similar to Fig.~\ref{fig:lnL}, each point represents one galaxy cluster, colour coded according to the virial ratio, $\eta$. Given the strong dependence of \deltaL on $\eta$, the distribution of $\eta$ in the $(M,c)$ plane is very similar to that of \deltaL, with variations mostly along the positive correlation direction of $M$ and $c$. For convenience, hereafter we call the negative correlation direction in the halo mass and concentration distribution as the ``degeneracy" direction, and the orthogonal direction to it as the ``biased" direction. 

The degeneracy in the best-fitting mass and concentration parameters reflects that dynamical models mostly constrain the gravity or mass at the median radius of the tracer, $M(r_{1/2})$, but are not sensitive to the shape of the mass profile (\citealp{Han2016a}. As a result, fits with similar level of biases at the tracer median radius, $b_{1/2}\equiv M_{\rm fit}(r_{1/2})/M_{\rm true}(r_{1/2})$, also have similar \deltaL, as shown in the right panel of Fig.~\ref{fig:2fea}. Comparing the two panels, it becomes obvious that the information carried by $\eta$ on the \DS can be equivalently summarized as the half-mass bias information. 

\begin{figure}
    \centering
    \includegraphics[width=0.45\textwidth]{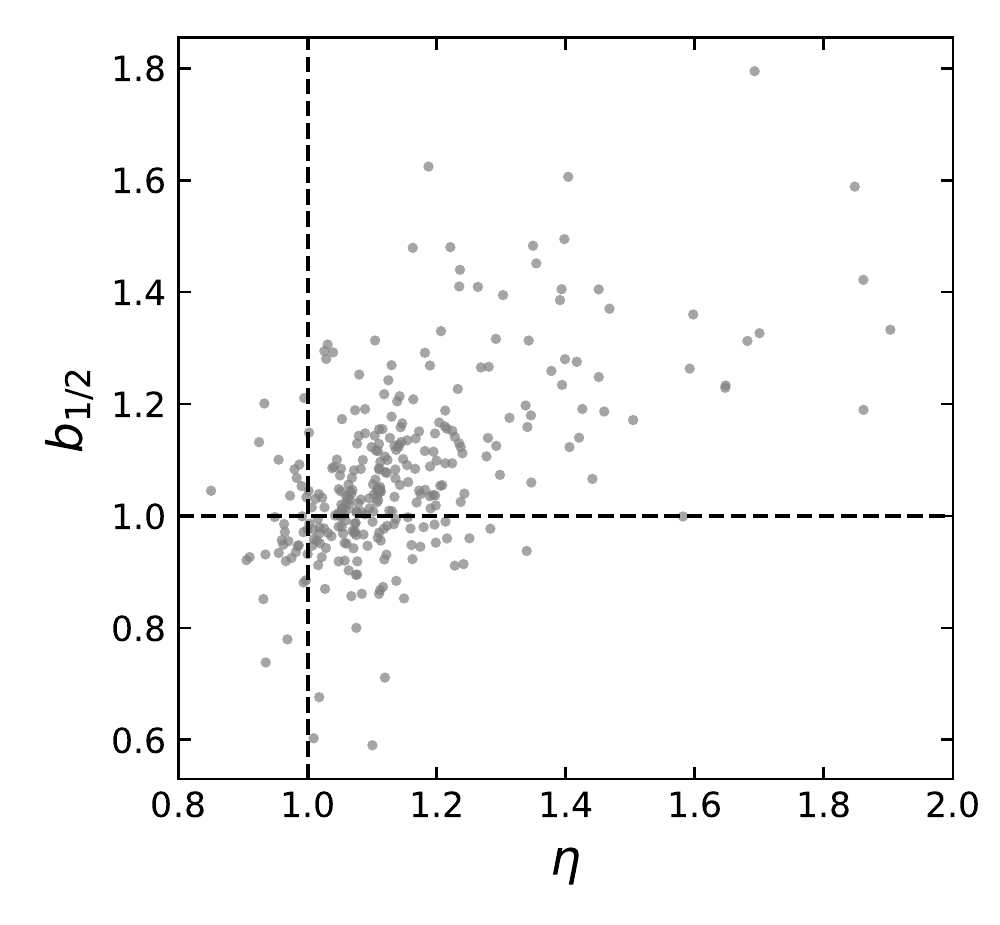}
    \caption{The half-mass bias from the fitting versus the virial ratio for the halos in our cluster sample.
    }
    \label{fig:b_eta}
\end{figure}

This connection between $\eta$ and $b_{1/2}$ is not difficult to understand considering the steady-state or equilibrium assumption of dynamical modelling. Clusters with the highest $\eta$ values are usually unrelaxed systems with high kinetic energy, which may be caused by major mergers or active mass accretion, and thus the kinetic energy is increased within a short time. If one still assumes the system is in equilibrium, the underlying gravity will be overestimated according to the virial theorem, leading to a large and positive $b_{1/2}$. Similarly, a low $\eta$ value would result in a small or negative $b_{1/2}$. The correlation between $\eta$ and $b_{1/2}$ is shown more explicitly in Fig.~\ref{fig:b_eta}. Systems with $\eta\approx 1$ are on average unbiased with $b_{1/2}\approx1$, while large $\eta$ halos typically have large $b_{1/2}$.

As we have pointed out before, clusters with the largest biases are not symmetrically distributed along the biased direction. This can be seen as the highest $\eta$ clusters are mainly distributed in the upper right corner of Fig.~\ref{fig:2fea}, with a large $b_{1/2}$. On the other hand, there is a shortage of points in the lower left corner, corresponding to a lack of cold but over-compressed systems. This is also obvious according to the distribution in Fig.~\ref{fig:b_eta}. This leads to an important conclusion that if no selection on cluster dynamical state is made, the estimated mass profiles of cluster halos will tend to be biased towards positive values of $b_{1/2}$ on average. 

Despite the fact that $M(r_{1/2})$ can be constrained the best than the mass within other radii \citep[e.g.][]{Wolf2010,Walker2011,Han2016a}, the associated bias, $b_{1/2}$, may still be significant for clusters which are strongly out of equilibrium. Thus selecting systems which are close to be in steady states is important, and our study in this paper provides critical clues of how to quantify the \DS of galaxy clusters. Theoretically, $\eta$ is the best individual feature out of our feature pool. In addition to $\eta$, we have explicitly examined other features with high importances, including $\chi_{\rm DS}, \dot{M}_\mathrm{a}$ and $f_{\rm ste}$. We find similar but weaker gradients in these features along the biased direction. 

In a few previous studies, no significant connections have been made between features of galaxy clusters and the bias in the best constrained virial mass. For example, \citet{Cialone2018} showed that the morphological parameter, $\mathcal{M}$, does not show strong connections with the bias in mass in their simulations. \citet{Gianfagna2021} also reported no significant dependence of the mass bias on $f_{\rm s}$ and $\Delta_{\rm r}$. Indeed, $\mathcal{M}$, $f_{\rm s}$ and $\Delta_{\rm r}$ all have lower \OOB scores than that of $\eta$ (Fig.~\ref{fig:one}).



\section{Conclusions} \label{sec:con}

In this work, we investigate the importances of a large variety of features in determining the \DS of galaxy clusters using RFR. We have 310 galaxy clusters in total, selected from the $z=0$ snapshot of the Three Hundred project, which also have associated mock optical, X-ray and SZ maps. The target variable is the log-likelihood ratio between the best-fitting and true halo parameters based on the \oPDF method, \deltaL, which quantifies the amount of systematics in the dynamical fit to the halo. The features of galaxy clusters are either constructed from the simulation in 3-dimension or from the corresponding optical, X-ray and SZ mock maps in projection. 
We used the \OOB score of each feature to evaluate its importance. 

We first present the overall RFR performance using features from separate categories and their combinations. Then, we focus on the importance of individual features and investigate the correlations among different features. We also explore the best feature combinations in determining the \DS of galaxy clusters. Based on these analysis, we find the virial ratio, $\eta$, to be the most important and independent feature in describing the \DS of clusters. 

Our main conclusions are summarized as follows:  

\begin{itemize}

\item Even after using all available features, the \DS of galaxy clusters cannot be fully captured. The maximum amount of predictable variation in \deltaL by our model only reaches $\sim 40\%$. This means our feature set have not exhausted the dynamical information of the cluster, despite that we have used an extensive set. By contrast, when using the performance of an alternative dynamical model, the SJE $\chi^2$ as a feature, we are able to predict the variation in \deltaL up to $\sim 80\%$, with the remaining part unpredictable due to statistical fluctuations in the target.

\item Features directly constructed from the simulation in 3-dimensions may contain more information about the \DS than those from mock maps in X-ray or SZ. 

\item Features defined through position and position offsets in the $D_2$ category are more important than the ICM morphological parameters from the X-ray and SZ maps.

\item The virial ratio, $\eta$, is the most independent and important single feature.  The good performance of $\eta$ can be tracked down to and understood as its significant correlation with the mass bias at the median tracer radius, $b_{1/2}$.

\item In addition to $\eta$, the fraction of stellar mass in the BCG and ICL, $f_{\rm ste}$, the age-weighted mass accretion rate, $\dot{M}_\mathrm{a}$, and the few relaxation parameters (e.g., $\chi_{\DS}$, $f_{\rm s}$ and $\Delta_{\rm r}$) are also important.

\item A few physical features, such as the SFR, age, spin, temperature and properties related to the substructures are poor indicators to the \DS. 

\item The same ICM morphological parameters can have very different importances between X-ray and SZ maps. 

\item For features defined through position and position offsets, the intensity weighted centroid positions based on the X-ray and SZ maps ($R_{\rm X}$ and $R_{\rm SZ}$) are more significant, whereas the intensity weighted centroid position of the BCG based on the optical maps, $R_{\rm BCG}$, is less important.

\item The minor to major axis ratio, $c/a$, has significantly lower importance than those features implying the current relaxation status (e.g., $\eta$, $\chi_{\rm DS}$ and $\Delta_{\rm r}$). This indicates the deviation from spherical symmetry is a less important source of systematic in dynamical modelling than the deviation from steady state.

\item By investigating the \OOB scores with different numbers of feature combinations, we find a combination of up to three different features from the $D_3$ category or from the position and position offset related features in the $D_2$ category, can already capture most of the information available to all the investigated features about the \DS, with an \OOB score reaching $\sim 0.4$. On the other hand, morphological features measured from X-ray and SZ maps are less important at any number of feature combinations.

\item Based on the analysis of Pearson correlation coefficients, we divide the features into four classes: i) features quantifying the current relaxation status (e.g., $\eta$ and $\chi_{\rm DS}$); ii) features quantifying the merger history (e.g., $\dot{M}_\mathrm{a}$ and $f_{\rm ste}$); iii) properties of satellites/subhaloes (e.g., $r2$); and iv) properties of the ICM (e.g., $T_{500} $ and $n_{\rm e}$). The \DS of galaxy clusters can be well captured with a combination of three different types of features listed above. 

\item If no selection to dynamical states are made, dynamical modelling of cluster halos tend to be biased towards a higher mass at the tracer median radius on average. This is due to the existence of a population of dynamically hot halos experiencing recent major mergers.

\end{itemize}

Our work provides an important reference on how to discriminate galaxy clusters of varying levels of equilibrium in numerical simulations and in real observations. 
Based on our conclusions above, theoretically, we recommend including $\eta$ as the most important feature upon determining the \DS of galaxy clusters, and if in combination with one or two more features related to the merger history, to the satellite and ICM properties, the \DS can be even better indicated. We suggest taking a priority of the features containing the information of the present relaxation status of galaxy clusters and halo assembly histories. Observationally, we recommend the usage of the offsets between the intensity weighted centroid positions on the X-ray and SZ maps and the intensity weighted BCG positions on the optical map, or the offsets between the intensity weighted X-ray/SZ positions and the position of the peak intensity pixel. The larger the offsets, the more unrelaxed the systems are. With good quantifications of the cluster \DS, it might be possible to correct for the bias in the best constrained mass profiles in dynamical modelling.

\section*{Acknowledgements}
This work is supported by NSFC (11973032, 11833005, 11890691, 11890692, 11621303, 12022307),
National Key Basic Research and Development Program of China
(No.2018YFA0404504), 111 project No. B20019 and Shanghai Natural Science Foundation, grant Nos. 15ZR1446700, 19ZR1466800. We acknowledge the science research grants from the China Manned Space Project with No. CMS-CSST-2021-A02, CMS-CSST-2021-A03 and CMS-CSST-2021-B03. We thank the sponsorship from Yangyang Development Fund.

This work has been made possible by the ‘The Three Hundred’ collaboration. The project has received financial support from the European Union’s H2020 Marie Skłodowska-Curie Actions grant
number 734374, i.e. the LACEGAL project. 
The simulations used in this paper have been performed in the MareNostrum Supercomputer at the Barcelona Supercomputing Center, thanks to CPU time granted by the Red Espa$\tilde{\rm n}$ola de Supercomputaci$\acute{\rm o}$n. The CosmoSim database used in this paper is a service by the Leibniz-Institute for Astrophysics Potsdam (AIP). The MultiDark database was developed in cooperation with the Spanish MultiDark Consolider Project CSD2009-00064.

This work has made extensive use of the \textsc{python} packages -- \textsc{ipython} with its \textsc{jupyter} notebook \citep{ipython}, \textsc{numpy} \citep{NumPy} and \textsc{scipy} \citep{Scipya,Scipyb}. All the figures in this paper are plotted using the python matplotlib package \citep{Matplotlib}. This research has made use of NASA's Astrophysics Data System and the arXiv preprint server. The computation of this work is partly carried out on the \textsc{Gravity} supercomputer at the Department of Astronomy, Shanghai Jiao Tong University.

We are very grateful for useful discussions with Prof. Zhiyuan Li from Nanjing University. 

\section*{Data Availability}
The data underlying this paper will be shared on reasonable request to the corresponding author.



\bibliographystyle{mnras}
\bibliography{paper} 




\appendix
\bsp	
\label{lastpage}
\end{document}